\documentclass[aps,reprint,superscriptaddress,preprintnumbers]{revtex4-1}
\usepackage{amsmath}
\usepackage{xr-hyper}
\usepackage{hyperref}
\usepackage{isotope}
\usepackage{mhchem}
\usepackage{xcolor}
\usepackage{graphicx}
\usepackage{mathrsfs}
\usepackage{dcolumn}
\usepackage{bm}
\usepackage{cases}
\usepackage{multirow,booktabs}
\usepackage{makecell}
\usepackage{amsmath}
\usepackage{amssymb}
\usepackage{sidecap}
\usepackage{graphicx}
\usepackage{here}

\usepackage{adjustbox}

\usepackage{tikz}
\usepackage{float}
\usepackage{tabularx}
\usepackage{soul}
\usepackage[normalem]{ulem}

\usepackage{xcolor}
\definecolor{pastelgray}{rgb}{0.81, 0.81, 0.77}
\definecolor{beaublue}{rgb}{0.9, 0.9, 0.93}

\newcommand{\cgmf} {$\texttt{CGMF}$}

\usepackage{todonotes}

\begin{document}
\title{Deblurring fission fragment mass distributions}
\author{Pierre Nzabahimana}
\email{pierre@lanl.gov}
\affiliation{Los Alamos National Laboratory, Los Alamos, NM, 87545, USA}
\author{Amy E. Lovell}
\affiliation{Los Alamos National Laboratory, Los Alamos, NM, 87545, USA}
\author{Patrick Talou}
\affiliation{Los Alamos National Laboratory, Los Alamos, NM, 87545, USA}
\affiliation{Stardust Science Labs, Santa Fe, NM 87507, USA}
\preprint{LA-UR-25-21699}

\begin{abstract}
Measurements of fission fragment mass distributions provide valuable insights into the properties of fissioning systems and the dynamics of the fission process. Pre-neutron emission distributions, essential for fission fragment evaporation codes like \cgmf{}, are extracted from distributions that are always measured after neutron emission, as the time scale of the emission of prompt fission neutrons is too short for direct measurement before the emission. However, obtaining accurate pre-neutron emission distributions requires methods that eliminate the effects of mass resolution and detector efficiency.
We propose a deblurring technique based on the Richardson-Lucy (RL) algorithm, commonly used in optics for image restoration, to correct for these experimental effects. The RL algorithm uses the measured mass distributions and a transfer matrix to perform iterative deconvolution. The advantage of this method over others is that it does not assume any predefined shape such as a sum of Gaussians, as in \cgmf{}, for the distributions.
In this paper, we apply the algorithm to the fission fragment mass distributions measured in the spontaneous fission of $^{252}$Cf to extract pre-neutron emission fission fragment mass distributions. The results from deblurring are then used as inputs to \cgmf{}, and we compare the \cgmf{} results obtained using deblurring inputs with the default \cgmf{} results.
\end{abstract}
\maketitle
\section{Introduction}
Nuclear fission is crucial for many applications, including energy production, defense, non-proliferation, and nuclear medicine. Observables from fission  experiments, such as pre- and post-neutron emission fission fragment (FF) characteristics and prompt neutron and $\gamma$-ray correlations, provide valuable insights into the evolution of the fission process.
In particular, studying FF mass distributions reveals asymmetry patterns in the fragment distribution. When a nucleus undergoes asymmetric fission, it splits into two fragments of different masses, referred to as light and heavy fragments. This asymmetry arises from nuclear shell effects, which favor the formation of fragments near doubly magic nuclei, such as tin (Sn), which are relatively stable. The asymmetry in the fission fragment mass distribution manifests as a double-humped distribution, in which the left-hand hump corresponds to light fragments and the right-hand hump corresponds to heavy fragments. There are also, some systems, like certain Fm isotopes that undergo symmetric fission, where the system splits into two fragments with mass distribution centered around a mass equal to half of the system's total mass.

The FF mass distribution is an important quantity of fission reactions in many ways: its features, such as peak positions and widths, can impact prompt neutron and $\gamma$-ray observables such as neutron and $\gamma$-ray multiplicities, as well as their energy spectra~\cite{Observables}. In addition, the FF mass distribution can impact the correlations of neutron multiplicity and $\gamma$-ray multiplicity with fragment mass. 
 Moreover, the pre-neutron-emission fission mass distribution, $Y_{pre}(A)$, is a crucial input to many theoretical models such as \cgmf{}, a fission fragment decay code developed at Los Alamos National Laboratory~\cite{Talou} and others such as $\texttt{FREYA}$~\cite{FREYA}, $\texttt{BeoH}$ \cite{BeoH,Lovell2021}, $\texttt{GEF}$~\cite{GEF}. \cgmf{} uses a Monte Carlo implementation of the Hauser-Feshbach statistical theory of nuclear reactions to simulate the decay of fission fragments on an event-by-event basis~\cite{Hauser}. It takes pre-neutron emission FF properties such as mass, charge, total kinetic energy, spin and parity distributions as inputs. The outputs of \cgmf{}, the histories of each fission event, which can then be analyzed to extract average values, distributions, and correlations, are then compared with experimental measurements. It is important to ensure that the input feeds to \cgmf{} are accurate to get outputs that agree with the experimental measurements as well as to get predictive power and reliable insight into the underlying fission process.
 
However, determining $Y_{pre}(A)$ is challenging because it can not be measured directly, as the time scale for the emission of neutrons is extremely short (about $10^{-14}$ sec). Instead, $Y_{pre}(A)$ is reconstructed based on assumptions about fission production and prompt neutron emission from primary fragments, see e.g. \cite{Duke2015}. Also, at this time, theoretical predictions of those scission fragment distributions remain limited in scope and accuracy for most applications.

Several methods have been used to measure the FF mass distributions. In particular the double-energy ($2E$) method, more extensively and double-energy double-velocity ($2E-2v$), where the number 2 refers to the fact that you are measuring the two fragments simultaneously. Other methods, such as the (E-v) method records only one fragment at a time~\cite{GASTIS2022166853}. This is because they use thicker samples with a thick backing that maintains the target in place. The fragment going away from the backing is the only one that can escape and be recorded. The advantage of this method is that you can use much bigger and stronger targets, therefore accumulating more statistics. The $2E$ method involves inferring $Y_{pre}(A)$ from the measured energies and angles of the fission products using kinematic equations \cite{Duke2015}. This method is iterative and requires corrections for prompt neutron emissions and pulse height defects \cite{Duke2015}. It proceeds by adding the neutron multiplicity ($\overline{\nu^*}$) to the measured ($Y_{post}$), following the equation $Y_{pre}(A) = Y_{post}(A) + \overline{\nu^*}(A)$. 
Alternatively, the $2E-2v$ method can be used. The method has led to the development of several instruments worldwide, such as SPIDER~\cite{MEIERBACHTOL201559}, VERDI~\cite{FREGEAU201635}, and COSI-FAN-TUTTE \cite{doi:10.1080/00337578608208713}. In the $2E-2v$ method, both the energies and velocities of the fission fragments are measured simultaneously and using the four measured quantities, both $Y_{post}(A)$ and $Y_{pre}(A)$ can be calculated \cite{refId0}. This method requires the measurement of both fragments simultaneously, which implies the use of thin targets that significantly limit the achievable statistics. So far, mostly measurements on $^{252}$Cf(sf) have been attempted for this reason. Thus, those assumptions in the methods, along with the mass resolution in the FF mass distribution measurement that smears the true distribution, can impact the resulting $Y_{pre}(A)$.  
In this paper, we propose the use of a deblurring approach to extract $Y_{pre}(A)$ from $Y_{post}(A)$ as well as removing the effect of mass resolution in the experimental $Y_{pre}(A)$ data. The procedure employs the Richardson-Lucy (RL) algorithm, initially developed to restore blurred images in optics~\cite{richardson1972bayesian, lucy1974iterative}. The derivation of the RL algorithm is based on Bayes' theorem~\cite{Phillips_2021} and follows an iterative procedure to find a self-consistent solution using only the measured distribution and the discretized response matrix of the apparatus, also known as the Transfer Matrix (TM). 

Recently, the technique has been successfully applied to various problems in nuclear physics such as the restoration of the decay energy spectrum of $^{26}\rm{O}\rightarrow ^{24}\rm{O}+n+n$ measured via invariant mass spectroscopy by the MoNa collaboration at National Superconducting Cyclotron Laboratory (NSCL)~\cite{PhysRevC.107.064315} and as an inverse problem method, see Refs.~\cite{NZABAHIMANA2023138247,tam2025source,nzabahimana2025source,chinesePL}. In the latter, the deblurring method was used to determine source function from two-particle correlations in heavy ion collisions. The source function is used as a tools to assess space-time characteristics of particle emission in heavy-ion collisions. Other applications of the method to nuclear physics problems can be found in Refs.~\cite{XU2024139009,danielewicz2022deblurring,pawelSideridge}.
In the case of inverse problem, the deblurring process progresses without directly inverting TM, which is uncommon in solving inverse problems~\cite{grech2008review}. By maintaining the positivity of the restored spectrum throughout the iterative procedure and avoiding direct TM inversion, the method circumvents severe singularity issues that typically arise in inverse problems.

In this paper, we apply the deblurring method to FF mass distributions measured in the spontaneous fission of $^{252}\text{Cf}$~\cite{HAMBSCH1997347,Romano,MEIERBACHTOL201559,knitter}. This reaction is widely used as a {\it standard} for studying nuclear fission, and extensive experimental data are available and have been analyzed for this system. The distribution obtained using the deblurring method represents, ideally, the "true" distribution, $Y_{pre} (A)$, in which the mass resolution and detector response biases have been removed.
We then use this distribution as input for \cgmf{} and compare the results with those obtained using the default \cgmf{} inputs, keeping all other parameters unchanged.

The rest of this manuscript is organized as follows: In Sec.~\ref{sec.2}, we discuss the theoretical models implemented in \cgmf{} that we use in this study. Section~\ref{sec:deblurring} discusses the deblurring method results.  We dedicate Sec.~\ref{sec:observables} to \cgmf{} results. Finally, conclusions and outlook are presented in Sec.~\ref{sec:summary}.

\section{Theoretical models \label{sec.2}}

\subsection{Default \cgmf{} FF mass distribution input}
Several efforts have been made to develop models and codes for analyzing nuclear fission measurements, particularly for neutron and photon emissions~\cite{Talou,FREYA,GEF,BeoH,Lovell2021}. Two types of models, deterministic and stochastic, are commonly used for nuclear fission analyses. Deterministic models, such as the Hauser-Feshbach fission fragment decay code $\texttt{BeoH}$ \cite{BeoH,Lovell2021}, are fully determined by initial conditions and input parameters. Those are beyond the scope of this paper and will not be discussed further.
In contrast, stochastic models introduce randomness, meaning that identical inputs and initial conditions do not guarantee the same outcomes. These models rely on Monte Carlo simulation methods, where the results, when averaged over many events, converge to those from deterministic approaches, assuming the same underlying models.

Here, we focus on the Monte Carlo FF decay code, \cgmf{}, which has the advantage over deterministic ones of being able to track the histories of neutron and photon emissions, allowing for the determination of correlations between emissions. 
Below, we give a brief outline of \cgmf{} and then refer readers to Ref.~\cite{Talou} for more details. For each fission event, the output is recorded in the history file and can be statistically analyzed to be compared to experimental data.
\cgmf{} uses a Monte Carlo implementation of the Hauser-Feshbach statistical theory~\cite{Hauser,Hauser-F, HFmodel} to follow the deexcitation of pre-neutron emission fission fragments. Since those fragments are neutron rich, the Hauser-Feshbach statistical model describes their decay through neutron and $\gamma$-ray emission to the ground state.
For a given fission reaction, the \cgmf{} code samples from a pre-neutron FF distribution as a function of mass, charge, total kinetic energy, spin, and parity, $Y(A,Z,\mathrm{TKE},J,\pi)$.


 The total kinetic energy (TKE) distribution for each mass split is represented by a Gaussian, with the mean and variance fitted to experimental data but scaled to reproduce the average TKE for a given incident neutron energy. Once the TKE is sampled, the total excitation energy (TXE) for the FF pair is determined through energy conservation, where $\mathrm{TXE} = Q - \mathrm{TKE}$, with $Q$ being the $Q$-value of the reaction. Subsequently, this provides insight into prompt fission neutron and $\gamma$-ray observables, since TXE is the sum of the total $\gamma$-ray and neutron emission energies.  The TXE is shared between the two FF based on a ratio of temperatures, and the TKE is shared based on kinematics.
 

In \cgmf{}, the spin and party distribution for each FF follows a Gaussian distribution with positive and negative party chosen to be equally probable, 
$\rho(J,\pi)=\frac{1}{2}(2J+1)\exp {\left[-\frac{J(J+1)}{2B^2(Z,A,T)}\right]}$, where $B^2$ is defined as a function of FF temperature, T, $B^2=\alpha \frac{\mathcal{I}_0(A,Z)T}{\hbar^2}$, and $\mathcal{I}_0 (Z,A)$ is the moment of inertia of the FF (A,Z) in its ground state, $\alpha$ is an adjustable parameter that is utilized to reproduce prompt $\gamma$-ray observables, see Ref.~\cite{IonelPhysRevC} and references within. 

Additionally, the charge distributions as a function of fragment mass and incident neutron energy are given according to Wahl’s systematics~\cite{wahl}. The means are determined by the unchanged charge distribution, corrected for the observed charge polarization. The widths depend on the fragment mass and are fitted to available experimental data as a function of energy.
 
Finally, turning our focus to the FF mass distribution, it is important to note that \cgmf{} implements fragment mass distribution as a parameterized three-Gaussian model:
\begin{eqnarray}
    Y(A)=G_0(A)+G_1(A)+G_2(A),
    \label{3Gaussian}
\end{eqnarray}
where the symmetric fission mode is given by
\begin{eqnarray}
    G_0(A)&=&\frac{W_0}{\sigma_0 \sqrt{2\pi}}\exp{\left(-\frac{(A-\overline{A})^2}{2\sigma_0^2}\right)}.
    \label{system}
\end{eqnarray}
The parameters $\sigma_0$, $\overline{A} = A_p/2$, and $W_0$ are, respectively, the width, mean, and weight of the symmetric mode, where $A_p$ is the mass of the fissioning nucleus. The second and third terms of Eq.~\eqref{3Gaussian} represent the two asymmetric fission modes and are given by 

\begin{eqnarray} 
G_{1,2}(A) &=& \frac{W_{1,2}}{\sigma_{1,2} \sqrt{2\pi}} \Bigg[\exp\left(-\frac{(A - \mu_{1,2})^2}{2 \sigma_{1,2}^2}\right) + \nonumber\\
& & \exp\left(-\frac{(A - (A_p - \mu_{1,2}))^2}{2 \sigma_{1,2}^2}\right) \Bigg], 
\end{eqnarray}
where $\sigma_{1,2}$, $\mu_{1,2}$, and $W_{1,2}$ are, respectively, the widths, means, and weights of the Gaussian model for the asymmetric modes. In the case of neutron-induced fission, these parameters are function of the neutron incident energy (see Ref.~\cite{Talou} for details). \cgmf{} requires the parameters of the three-Gaussian model as input. These parameters are typically fit to experimental extractions of $Y_{pre}(A)$. 

\subsection{Deblurring framework}


The blurring formula between a measured function, $g$, and the true function, $\mathcal{F}$, is defined as
\begin{eqnarray}
g(x^{\prime})=\int dx \, M(x^{\prime},x) \, \mathcal{F}(x) \,  ,
\label{blur}
\end{eqnarray}
where, $x'$ is the measured variable, $x$ is the true variable, and $M(x^{\prime}, x)$ is the conditional probability that the particles emitted at $x$ are measured at $x'$, also known as response function or TM.
For example, in the optical deblurring problem, a photon is measured with a property $x'$, while its true property is $x$. 
When the properties are discretized, such as in attributing the photon to a particular pixel, Eq.~\eqref{blur} becomes in the matrix form between the distribution vectors:
\begin{equation}
g_i=\sum_j M_{ij} \, \mathcal{F}_j \,  .
\label{Eqq1}
\end{equation}

A deblurring method, such as RL, seeks to determine the distribution $\mathcal{F}$ given $g$ and $M$. To arrive at the RL strategy, a backward relation between $g$ and $\mathcal{F}$ is invoked, involving a conditional probability $P$ that complements $M$. By requiring the fulfillment of a Bayesian relation involving $M$ and $P$ ( more details about this relation can be found in Ref.~\cite{nzabahimana2023particle}), $\mathcal{F}$ can be found iteratively
\begin{equation}
    \mathcal{F}^{(\mathfrak{r}+1)} \equiv \mathcal{F}^{(\mathfrak{r})} \, A^{(\mathfrak{r})} \, ,
    \label{eq:RL}
\end{equation}
where the amplification factor  $A^{(\mathfrak{r})}$ is given as:
\begin{equation}
    A^{(\mathfrak{r})} =  \frac{\sum_i \alpha_i \frac{g_i}{g_i^{(\mathfrak{r})}} \, M_{ji} }{\sum_i \alpha_i M_{ji}} .
    \label{eq:RL0}
\end{equation}
Here, $\mathfrak{r}$ is the iteration index, $\alpha$ is a weight that specifies the relative importance of the given data in inferring $\mathcal{F}$ and $g^{(\mathfrak{r})}$ is the $\mathfrak{r}^{th}$ estimate of the measured distribution
\begin{equation}
    g^{(\mathfrak{r})}_i=\sum_j M_{ij} \, \mathcal{F}^{(\mathfrak{r})}_j \,  .
    \label{RLr}
\end{equation}
In this study, we implement RL algorithm using a Gaussian-shaped transfer matrix, 
 \begin{eqnarray}
     M(x|x^{\prime})\propto \frac{1}{\sqrt{2\pi\sigma_{TM}^2}}\exp\Big[-0.5\Big (\frac{x-x^\prime}{\sigma_{TM}}\Big )^2\Big ],
     \label{TM}
 \end{eqnarray} 
 where $\sigma_{TM}$ is the standard deviation.  Note that we do not expect the TM to be always Gaussian, but other forms can also be used, especially when the experimental TM is available, we recommend it to be used instead. Also, it is important to note that any information about the experimental setup or corrections related to the assumption made during data analysis can be included in the expression. For instance, in our case of FF mass distribution, we apply the correction about neutron emission when the RL algorithm is applied to measured post-neutron emission FF mass distribution to extract pre-neutron emission FF mass distribution (we discuss this in the next section).

 The success of the RL algorithm has been shown to rely on the assumption that the quantities $g$, $\mathcal{F}$, and $M$ are positively defined~\cite{lucy1974iterative}. The iteration starts with an initial guess of $\mathcal{F}^0$, where any positive value can be used to initiate the guess. For example, references \cite{nzabahimana2023particle,PhysRevC.107.064315} use $\mathcal{F}^0 = 1$. However, the convergence rate of the solution depends on the choice of $\mathcal{F}^0$.

Despite its advantages such as ease of use, computational efficiency, and not assuming a specific shape for $Y_{pre}(A)$, the deblurring method has certain limitations. It utilizes measurements recorded as histograms, requiring careful rebinning of the TM to match binning in the data. Moreover, each experiment requires its own TM, which must be accurately modeled or simulated.
In addition, it is worth mentioning that as with other deconvoluting methods, the RL algorithm depends on the number of iterations and may suffer from instability after a few iterations due to noise amplification. This issue appears if the measured/blurred distributions, $g$, fluctuate a lot, for example, distributions measured with limited statistics. To overcome this issue, a regularization technique is needed.
Different levels of regularization can be applied to the RL algorithm depending on the statistics of the measured distributions. These include binning and applying a small regularization factor to Eq.~\eqref{eq:RL} to suppress instabilities during the iterations, as in Refs.~\cite{NZABAHIMANA2023138247,PhysRevC.107.064315,danielewicz2022deblurring}. In the latter, regularization techniques, such as the one known as Total Variation (TV), are commonly used \cite{PhysRevC.107.064315,NZABAHIMANA2023138247}. The TV regularization  is a denoising algorithm that uses nonlinear combinations of derivatives of the restored spectra, e.g. \cite{PhysRevC.107.064315} and references therein. In the case of low-statistics distributions with strong fluctuations, it is necessary to combine  binning and TV regularization. However, for high-statistics measured distributions and a low level of fluctuation, such FF mass distribution, binning regularization alone is sufficient to mitigate those instabilities. 

To validate the RL algorithm, we perform two tests examples using known input distributions. In each case, the original distribution is convolved with the TM, Eq.~\eqref{TM} and the RL algorithm is applied to the resulting blurred distribution; first in the absence of noise, and then with added noise, to assess the algorithm's ability to reconstruct the original distribution. These test cases are presented in Appendix~\ref{testsec}, and in all cases, the original distribution shape is accurately recovered.

\section{Deblurring results for fission fragment mass distributions}
\label{sec:deblurring}

In this section, we discuss the results obtained by applying the deblurring method to experimental fission fragment (FF) mass distributions reported for the spontaneous fission of $^{252}$Cf. Using this method, we extract the "true" pre-neutron distribution, $\tilde{Y}_{\text{pre}}(A)$, in two different ways: by removing the effects of mass resolution from the reported $Y_{\text{pre}}(A)$\cite{Romano,HAMBSCH1997347}, and by correcting for neutron emission effects in the measured $Y_{\text{post}}(A)$~\cite{Romano,MEIERBACHTOL201559}. The results for both cases are discussed in Subsections~\ref{ssectA} and~\ref{ssectB}, corresponding to the deblurring of $Y_{\text{pre}}(A)$ and $Y_{\text{post}}(A)$, respectively.

We can express the measured mass distribution in a discritized form the same way as in Eq.\eqref{Eqq1}, 
\begin{equation}
Y_i=\sum_i P_{ij} \, \tilde{Y}_j \, .
\label{YApost}
\end{equation}
We obtain Eq.~\eqref{YApost} by mapping the fission fragment mass distributions onto the blurring relation, which corresponds to substituting $ g_i$ by $Y(A) \equiv Y_i$, $\mathcal{F}_j$ by $\tilde{Y}(A) \equiv \tilde{Y}_j$ and $M(x|x^{\prime})\equiv M_{ij}$ by $P(A|A^{\prime}) \equiv P_{ij}$,
 For simplicity, we use \(\tilde{Y}(A)\) in Eq.~\eqref{YApost} to refers to the deblurred distribution, the estimate of the true pre-neutron emission FF mass distribution, the fragment mass distribution after the compound nucleus splits into two fragments but before the fragments emit any neutrons, and \(Y(A)\)  measured distribution (i.e., measured $Y_{pre}(A)$ or $Y_{post} (A)$). The same substitution is also applied to the expressions for RL algorithm, Eqs.~\eqref{eq:RL}-\eqref{RLr}.
The advantage of expressing \(Y\) as in Eq.~\eqref{YApost} is that it enables us to directly apply the RL algorithm, to solve for \(\tilde{Y} (A)\). 

Here, \(\mathbf{P}\) should be understood as conditional probability of detecting a fragment of mass $A$ when a fragment of mass $A^{\prime}$ was produced during the fission reaction. We implement RL algorithm using a Gaussian-shaped as in Eq.~\eqref{TM}, by simply replacing the variables $x$ and $x^{\prime}$ by $A$ and $A'$. The widths $\sigma_{TM}$ are provided in Table~\ref{tab:my_label} for the different experimental datasets. 

\subsection{Deblurring results for reported pre-neutron emission FF mass distribution} \label{ssectA}
\begin{figure}[H]
    \centering
    \includegraphics[width=.91\linewidth]{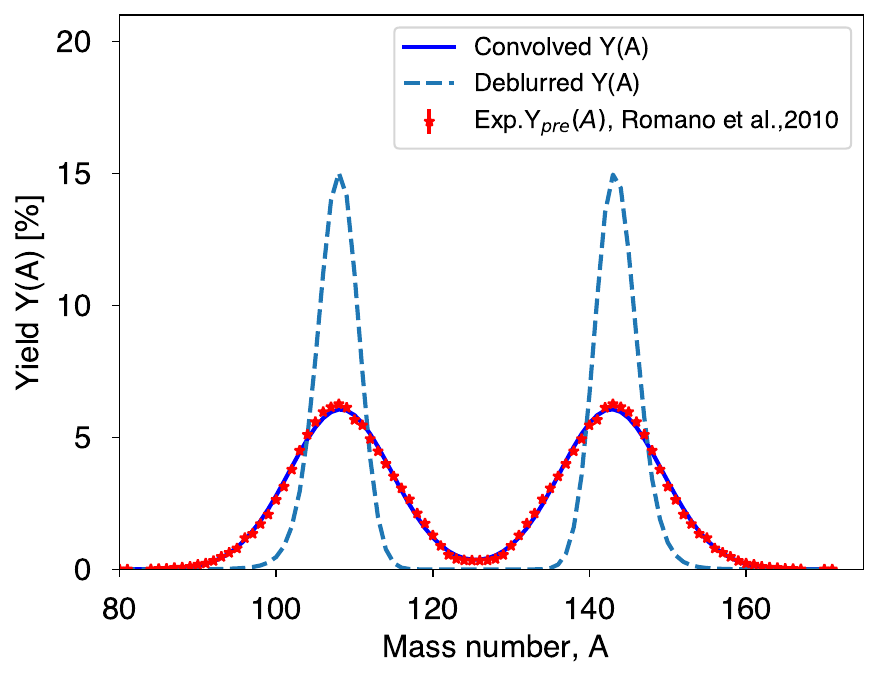}
    \caption{Deblurred $Y_{pre} (A)$, dashed line obtained by deblurring pre-neutron emission FF mass distributions (red stars) extracted from  Romano, {\it et al}~\cite{Romano} for $^{252}$Cf(sf). The solid line represents the convolved distribution obtained using Eq.~\eqref{RLr}, or Eq.~\eqref{YApost}, by convolving the deblurred distribution with the TM.
}
    \label{ratio}
\end{figure}

\begin{figure}[hpt]
    \centering
    \includegraphics[width=1\linewidth]{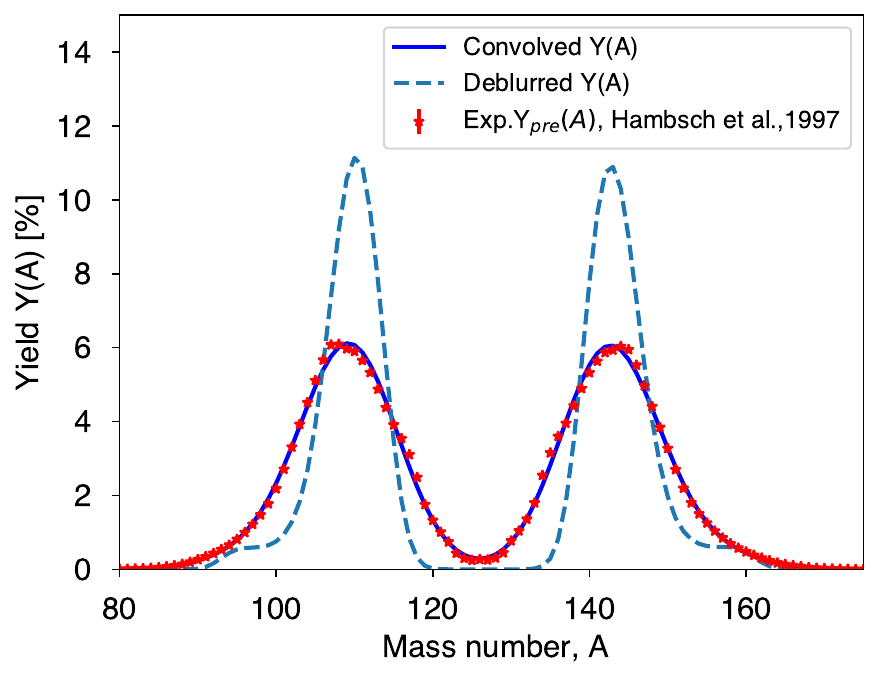}
    \caption{Same as Fig.~\ref{ratio} for pre-neutron emission FF mass distribution from Ref.~\cite{HAMBSCH1997347}.}
    \label{YA3}
\end{figure}

First, the RL algorithm is applied to measured $Y_{pre}(A)$.
In Figs.~\ref{ratio} and ~\ref{YA3}, we show the deblurred distributions obtained by applying the RL algorithm on experimental $Y_{pre}(A)$ data by Romano, \emph{et al.} \cite{Romano} and Hambsch, \emph{et al.} \cite{HAMBSCH1997347}, respectively.  The experimental $Y_{pre}(A)$ are plotted as the red stars, and the dashed lines represent \(\tilde{Y}_{pre} (A)\) obtained by using the RL algorithm in Eq.~\eqref{eq:RL}. Clearly, the difference between deblurred and experimental distributions is significant. For instance the peaks of \(\tilde{Y}_{pre} (A)\) are narrow and high compared to those of the experimental $Y_{pre}(A)$ curves. This is the evidence that the deblurring method has removed some experimental effects from the measured FF mass distribution. The half widths of \(\tilde{Y}_{pre} (A)\), compared to those of the measured distribution, are shown in Table~\ref{tab:my_label}. As can be seen in the table, in all cases, the width of the measured distribution is at least twice as large as that of the deblurred distribution.

Here, the iterative procedure begins with the measured distribution as the initial guess, $\tilde{Y}^0 = Y$, which leads to faster convergence. Ideally, the solution converges when the ratio $\frac{Y}{Y^{(\mathfrak{r})}} = 1$, i.e., when $Y^{(\mathfrak{r})} = Y$. In practice, due to noise in the measured distribution, exact agreement is not achievable. Instead, the algorithm seeks a solution $Y^{(\mathfrak{r})}$ that best reproduces the measured distribution.
At each iteration, starting from the initial guess $\tilde{Y}^0$, the estimate $Y^{(\mathfrak{r})}$ is computed by convolving the current solution $\tilde{Y}^{(\mathfrak{r})}$ with TM using Eq.~\eqref{RLr}. This procedure is repeated until iteration $\mathfrak{r}$, where the ratio $\frac{Y}{Y^{(\mathfrak{r})}}$ no longer improves.

\begin{figure}[H]
    \centering
    \includegraphics[width=.81\linewidth]{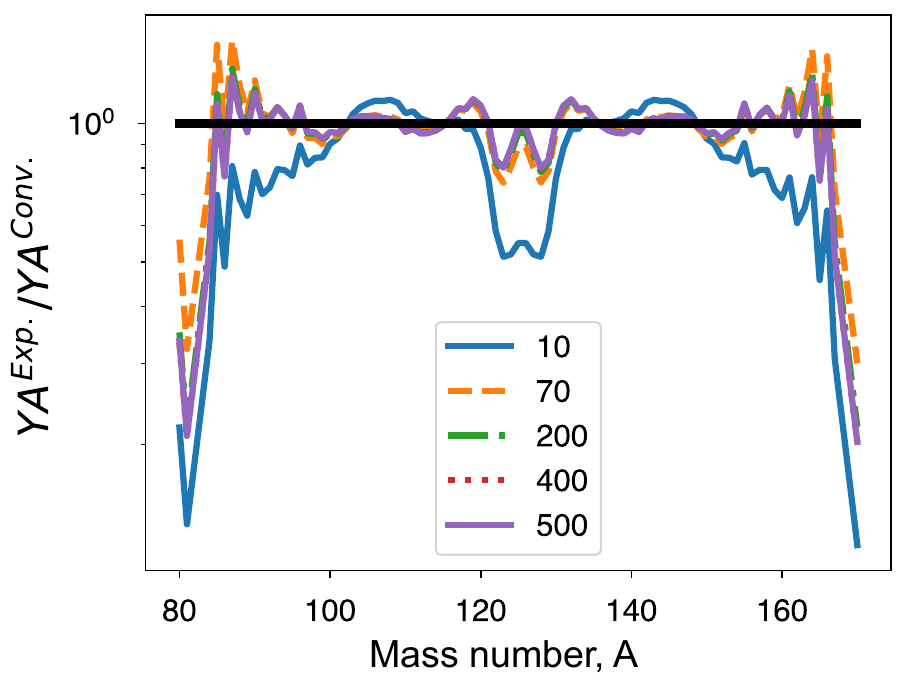}
    \caption{
    The ratio of experimental $Y_{pre}(A)$ (shown as red stars in Fig.~\ref{ratio})~\cite{Romano} to the convolved (blurred) distribution, in log scale as a function of fragment mass. Results are shown for a number of iterations, varying from 10 to 500.
}
    \label{ratio1}
\end{figure}

As an example, Fig.~\ref{ratio1} shows the ratio of the experimental FF mass distribution to the convolved FF mass distribution. The horizontal solid line represents the ideal ratio, which would result from a perfect match between the estimated and measured distributions. As noted earlier, when applying the RL algorithm to experimental data with noise, the goal is to stop iterating once the best estimate is reached. The different line styles, labeled by iteration number, indicate the ratio at various stages. After approximately 200 iterations, the ratio approaches unity and no longer improves.
In this case, we generally observe good agreement between the experimental and the FF mass distributions when $0.9 \leq \frac{1}{N} \sum_i \frac{Y_i}{Y^{(\mathfrak{r})}_i} \leq 1$, where $N$ is the number of data points in the distribution. This typically corresponds to $\mathfrak{r} \geq 70$ iterations.

It is important to note that we rebin the TM to align with the binning of the measured distribution. In this work, we use an $N \times N$ TM, where $N$ is the number of bins in the measured distribution. In some cases such as the inverse problem described in Ref.~\cite{NZABAHIMANA2023138247}, where the true and blurred distributions have different dimensions, a TM of size $N \times M$ is used. 

The deblurred distribution is compared to the experimental mass distribution by introducing a mass resolution function. This is achieved by convolving the deblurred distribution with the TM using Eq.~\eqref{YApost}. The resulting distribution, referred to as the convolved distribution, is shown as a solid line in Figs.~\ref{ratio} and ~\ref{YA3}. Note that when the TM used in the RL algorithm is a good approximation of the experimental response matrix and the deblurred distribution is the optimal solution, the convolved distribution should reproduce the experimental distribution. As shown in these figures, the agreement between the convolved distribution (solid line) and the experimental $Y_{pre}(A)$ (red stars) is good, which confirms that the deblurring has removed the mass resolution in the measurements.

\subsection{Deblurring results for measured post-neutron emission FF mass distribution}\label{ssectB}

Next, we apply the RL algorithm to estimate the true $Y_{\text{pre}}(A)$ from the measured $Y_{\text{post}}(A)$, effectively removing the impact of assumptions made in the experimental procedure. In this case, however, we introduce a parameter, ${\nu^*}$, into the TM to correct for neutron emission. 

With the inclusion of ${\nu^*}$, the TM described by Eq.~\eqref{TM} is modified such that the variable $A'$ becomes $ A' - \nu^*$. Since light and heavy fragments emit different numbers of neutrons, we introduce a fragment mass dependence for $\nu^*$, defined as ${\nu^*}(A) = mA + s$ within the TM. The parameters $m$ and $s$ are obtained by fitting to experimental average neutron multiplicity data. Note that those parameters are different for light and heavy fragment. Using the data from Ref.~\cite{zeynalov}, we obtain $m = 0.063$ and $s = -4.5$ for light fragments, and $m = 0.069$ and $s = -8.23$ for heavy fragments.
To validate our choice of $\nu^*$, we compare the resulting deblurred distribution to the three-Gaussian model of Eq.~\eqref{3Gaussian}. A good match with this model indicates a reasonable choice for $\nu^*$.
\begin{figure}[H]
    \centering
    \includegraphics[width=.91\linewidth]{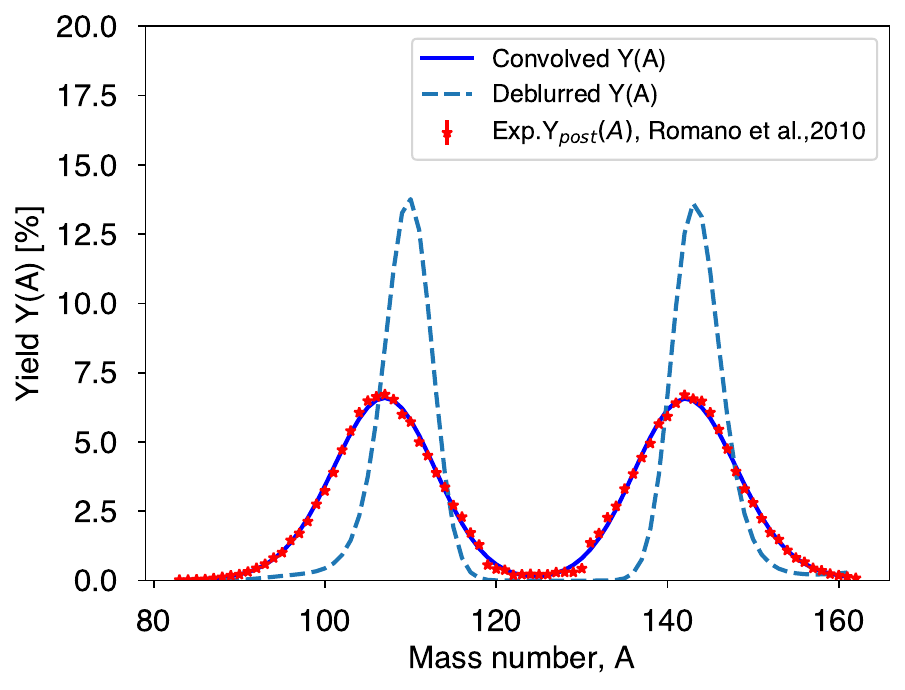}
    \caption{Fission fragment mass distributions for $^{252}$Cf(sf). Dashed line shows a FF mass distribution obtained from deblurring the experimental post-neutron emission FF mass distribution from Ref.~\cite{Romano} (shown by stars), using Eq.~\eqref{eq:RL}. Solid line shows  distributions results from convolving the deblurred distribution (dashed line) with TM. As observed in the figure, the  distribution reproduces the experimental data. The details are provided in the text.  
}
    \label{fig:YApost}
\end{figure}

\begin{figure}[hpt]
    \centering
    \includegraphics[width=1\linewidth]{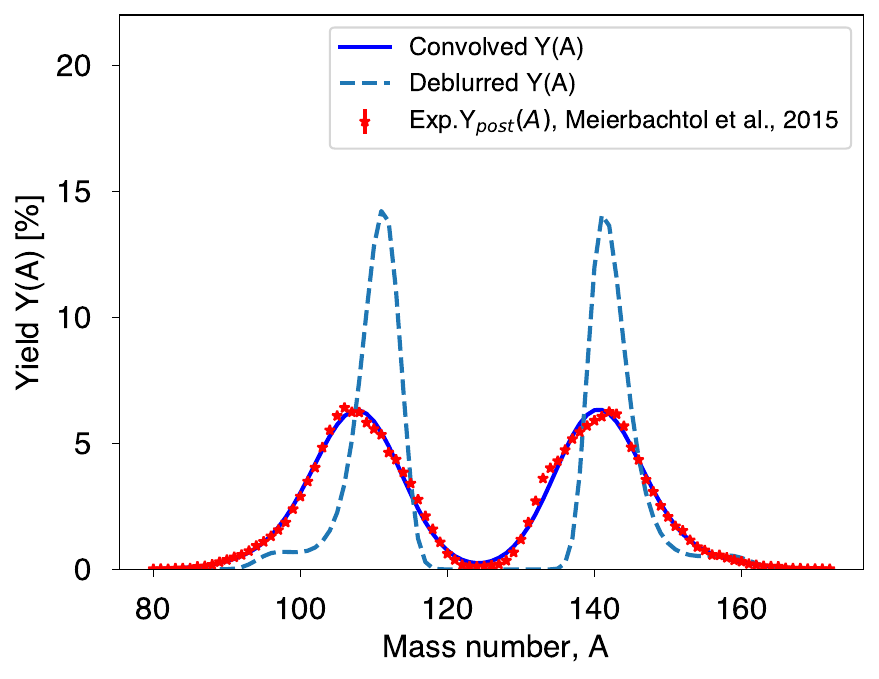}
    \caption{Same as Fig.~\ref{fig:YApost} but for post neutron emission FF mass experimental data reported in Ref.~\cite{MEIERBACHTOL201559}.}
    \label{YA2}
\end{figure}

\begin{figure*}
    \centering
    \includegraphics[width=0.6\linewidth]{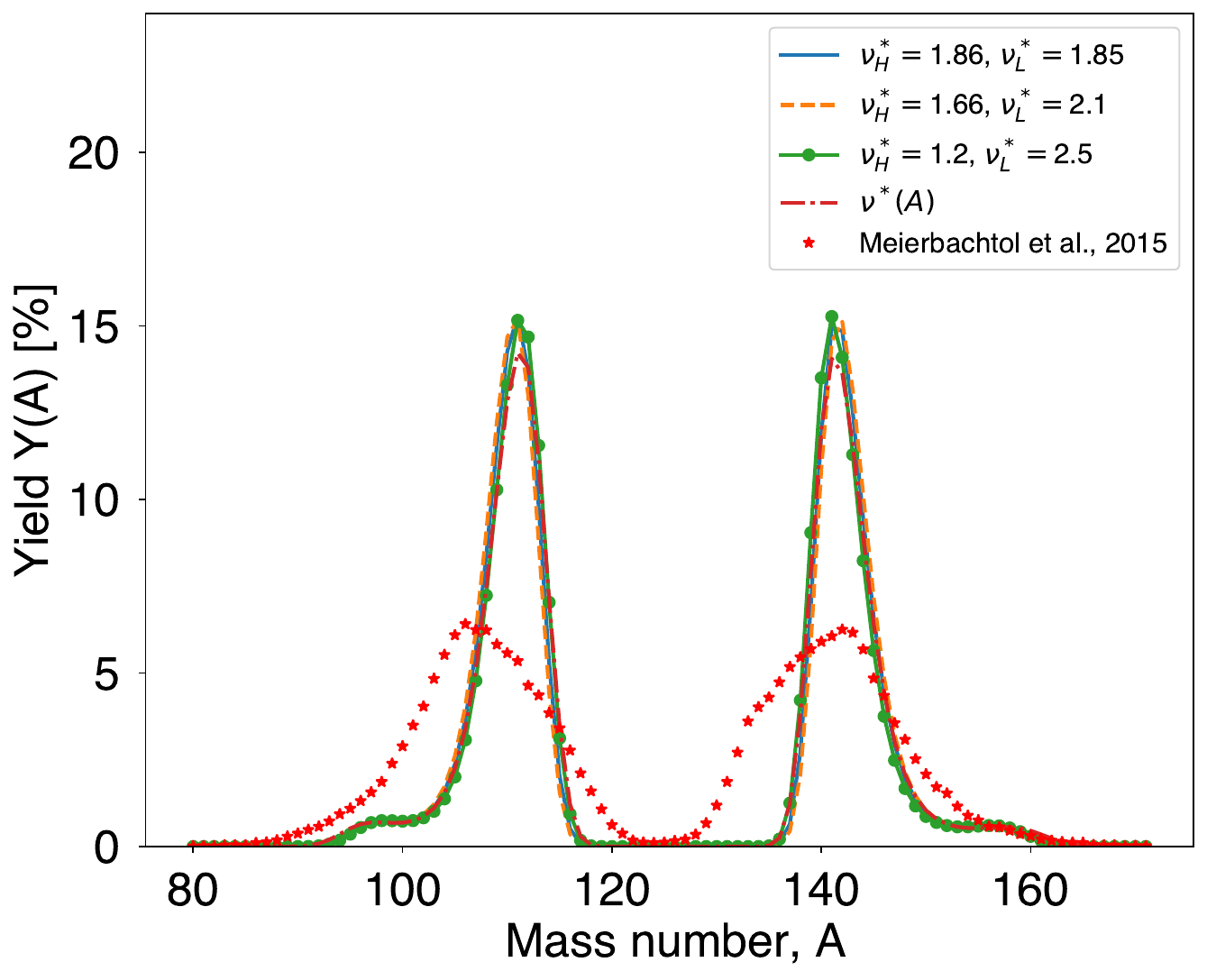}
    \caption{Comparison of deblurred $Y_{\text{pre}}(A)$ extracted from the measured $Y_{\text{post}}(A)$, taken from Ref.~\cite{MEIERBACHTOL201559}, using the Richardson-Lucy (RL) algorithm under different assumptions for $\nu^*$. The stars represent the experimental $Y_{\text{post}}(A)$, while the deblurred distributions are shown with different line styles corresponding to various values of $\nu^*_L$ (light fragment) and $\nu^*_H$ (heavy fragment). While overall differences are observed among the resulting deblurred distributions, they all exhibit similar general behavior.}
    \label{varynu}
\end{figure*}

In Figs. \ref{fig:YApost} and \ref{YA2}, the dashed lines represent the deblurred distributions obtained by applying the RL algorithm to the experimental $Y_{\text{post}}(A)$. The measured $Y_{\text{post}}(A)$ from $^{252}$Cf(sf), used to produce the results shown in both figures, were taken from Refs. \cite{HAMBSCH1997347} and \cite{Romano} and are shown as red stars.
These experimental $Y_{\text{post}}(A)$ distributions are processed using the RL algorithm to reconstruct the corresponding, \(\tilde{Y}_{pre} (A)\) that best estimates of the true pre-neutron fragment mass distributions. Similar to Figs. \ref{ratio} and \ref{YA3}, the distributions,\(\tilde{Y}_{pre} (A)\) are narrower and taller than the measured ones, due to the removal of mass resolution effects by the RL algorithm. The solid blue lines represent the convolved distributions, obtained in the same way as in Figs.~\ref{ratio} and \ref{YA3}. Again, these convolved distributions show good agreement with the experimental $Y_{\text{post}}(A)$.

Additionally, we observe a shift in the peak positions between the measured and deblurred $Y_{pre} (A)$. This shift corresponds to the average number of neutrons emitted per fragment and is introduced by the parameters $\nu^*$. Moreover, we observe asymmetry in the peak amplitude of the peak heights of the light and heavy fragments for the measured $Y_{post}(A)$, those asymmetries are corrected by including in the TM model a factor of 0.96 for Fig. \ref{fig:YApost} and 0.945 for Fig. \ref{YA2}.

\begin{table*}
   \caption{The first column, the references for the experimental data for $^{252} $Cf(sf) used in the RL algorithm. The second column shows the half-width values of the Gaussian model for TM ($\sigma_{TM}$). The third column lists the approximated values corresponding to the half-width of the peak in the measured distributions ($\sigma_{data}$), while the fourth column provides the values representing the half-width of the peak in the deblurred distributions ($\sigma_{debl.}$). }
    \label{tab:my_label}
    \centering
    
    \begin{tabular}{|l|c|c|r|}
    \hline
    Data Exp.data& $\sigma_{TM}$& $\sigma_{Data}$& $\sigma_{Debl.}$\\
    \hline
    $Y_{pre}(A)$ by Romano et al., 2010~\cite{Romano}, refer to Fig.~\ref{ratio} &6.2 &7.3  & 3 \\
    $Y_{pre}(A)$ by Hambsch et al., 1997~\cite{HAMBSCH1997347}, refer to Fig.~\ref{YA3}&5.3 &8 & 4\\
    $Y_{post}(A)$ by Romano et al., 2010~\cite{Romano}, refer to Fig.~\ref{fig:YApost}&5.2 &7 & 3.1\\
    $Y_{post} (A)$ by Meierbachtol et al., 2015~\cite{MEIERBACHTOL201559}, refer to Fig.~\ref{YA2}&5.2 &7.7 & 2.8\\
       \hline
    \end{tabular}
 
\end{table*}
Moreover, we assessed the impact of the choice of $\nu^{*}$ on the deblurred distributions by applying different prescriptions for $\nu^{*}$, including constant values (distinct for light and heavy fragments) as well as a mass-dependent form, $\nu^{*}(A)$ as illustrated in Figs.~\ref{fig:YApost} and \ref{YA2}. The resulting deblurred distributions, obtained using these various $\nu^{*}$ prescriptions, are shown in Fig.~\ref{varynu}.
As an example, we consider the measured $Y_{\text{post}}(A)$ distribution from Meierbachtol, \emph{et al.}~\cite{MEIERBACHTOL201559}. It is evident that the deblurred distributions vary depending on the chosen $\nu^*$, and we note that some of the $\nu^*$ values used in this comparison were chosen arbitrarily. While not all of these are physically realistic, they serve to illustrate the sensitivity of the deblurred results to the choice of $\nu^*$.
For the distributions shown in Fig.\ref{varynu}, we ensure that the sum of $\nu^*$ values for light and heavy fragments remains close to the standard average neutron multiplicity, $\overline{\nu} = 3.759 \pm 0.0048$, for $^{252}$Cf(sf)\cite{standard}. In this context, the assumed value of $\nu^*$ for the light (heavy) fragments represents the average neutron multiplicity associated with those fragments.

We observed a bumped structure around masses $A=$95 and $A=$155 in the deblurred results shown in Fig. \ref{YA3} and Fig. \ref{YA2}. However, we do not interpret these features to be associated with a specific fission mode, as there may be other experimental factors currently inaccessible to us that could be responsible for such structures. To evaluate the RL algorithm’s ability to restore low-amplitude peaks that are not visible in the blurred distribution, we performed a test described in Appendix~\ref{testsec}. For all levels of noise considered in the blurred data, the RL algorithm was able to successfully recover those peaks. Therefore, further investigation is needed to better understand the origin of the structures observed near these fragment masses.
\section{\cgmf{} calculations}
\label{sec:observables}



A key aspect of this study is the use of the pre-neutron FF distributions, obtained through the deblurring method discussed in the previous section, as input to the \cgmf{} code. As a baseline for comparison, we also perform \cgmf{} calculations with default input pre-neutron emission fission fragment yields. For default \cgmf{}, those parameters are obtained by fitting experimental $Y_{pre}(A)$ and for the deblurring method by fitting the deblurred distributions. The parameters are presented in Table~\ref{Table2}. The \cgmf{} numerical results discussed here were obtained with $3\times 10^5$ fission events. 

\begin{table*}
   \caption{ Parameters obtained by fitting the three-Gaussian model, Eq.~\eqref{3Gaussian}, to the deblurred Y(A) and the default \cgmf{} Y(A). We use these parameters in \cgmf{}'s three-Gaussian parametrization model for mass distribution inputs.}
    \label{Table2}
    \centering
    
    \begin{tabular}{|l|c|c|c|c|c|c|c|r|}
    \hline
    Experimental data&$W_1$&$W_2$&$\mu_1$&$\mu_2$&$\sigma_1$&$\sigma_2$ & $W_0$&$\sigma_0$ \\
    \hline
    Romano \emph{et al.,} 2010~\cite{Romano}, refer to Fig.~\ref{ratio} &0.3130&0.6457 & 108.59&145.09&1.966&2.78 & 0.0825&20\\
    Hambsch \emph{et al.,} 1997~\cite{HAMBSCH1997347}, refer to Fig.~\ref{YA3}& 0.911& 0.0808 &109.20&153.9 & 3.30&5.08& 0.0153 &20\\
    Romano \emph{et al.,} 2010~\cite{Romano}, refer to Fig.~\ref{fig:YApost}& 0.7284&0.0929&109&145.3 &2.1&3.57 &0.2251 &20\\
    Default \cgmf{}& 0.3562&0.6453&141.78&145.26 &5.914&7.906 &0.00 &10.2 \\
       \hline
    \end{tabular}
 
\end{table*}
\begin{figure}[hpt]
    \centering
    \includegraphics[width=1\linewidth]{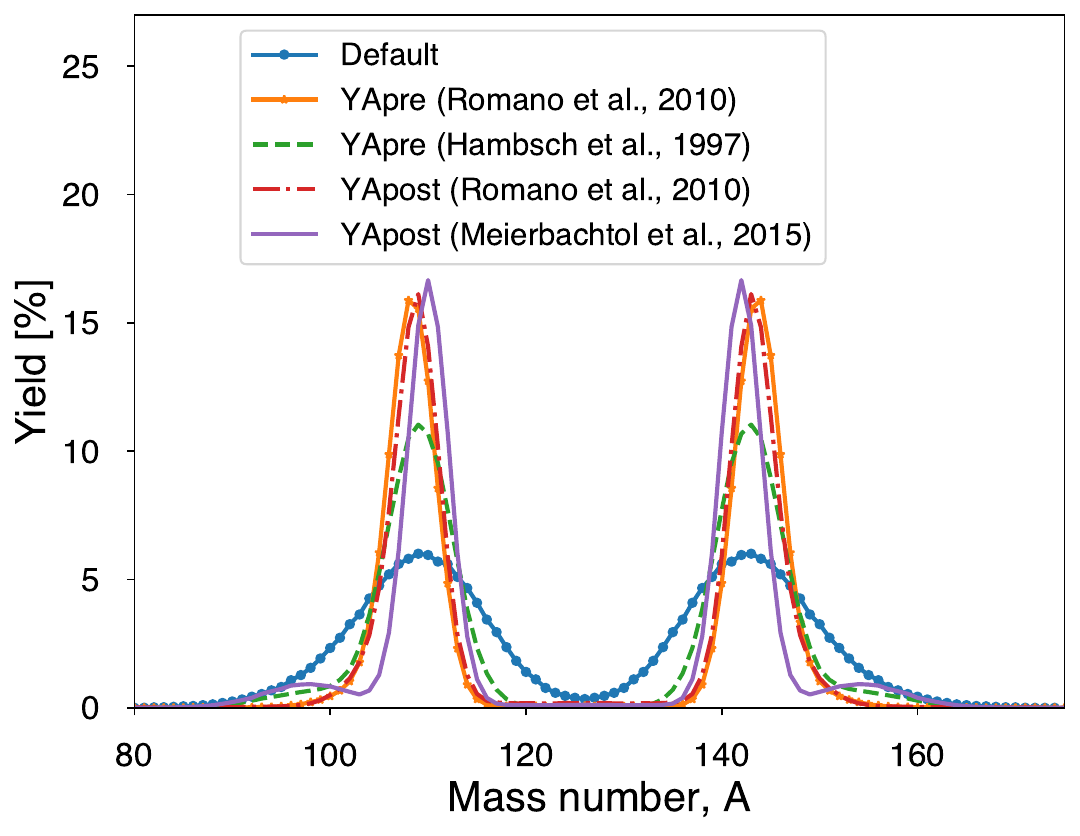}
    \caption{The default fission fragment mass distribution (solid blue line with dots) implemented in \cgmf{} for $^{252}$Cf(sf) is compared to FF mass distributions obtaining by deblurring experimental mass distributions: pre-neutron FF mass distribution  Romano, \emph{et al.}from~\cite{Romano} (solid line with stars ) and from Hambsch, \emph{et al.}~ \cite{HAMBSCH1997347} (dashed line green), and experimental post-neutron FF mass distribution from Romano, \emph{et al.}~\cite{Romano} (dot-dashed line red) and from Meierbachtol, \emph{et al.} ~\cite{MEIERBACHTOL201559} (purple solid line).}
    \label{CGMFdistribution}
\end{figure}
\subsection{Fission fragment mass distributions}
In this subsection, we discuss FF mass distribution calculation from \cgmf{} for $^{252}$Cf(sf) using different mass distribution inputs obtianed using deblurring. We use inputs obtained by deblurring both measured $Y_{pre}(A)$ and $Y_{post}(A)$, then the resulting \cgmf{} mass distributions are reblurred to be compared with default \cgmf{} mass distribution.

In Fig.~\ref{CGMFdistribution}, we compare the deblurred distributions shown in Figs.~\ref{ratio}--\ref{YA2} to the default distribution in \cgmf{}. The orange solid line with stars, the green dashed line,  the red dash-dotted line and the purple solid line represent the results obtained from deblurring the experimental pre-neutron fragment mass distribution reported by Romano, \emph{et al.} \cite{Romano} (see Fig.~\ref{ratio}), the experimental pre-neutron fragment mass distribution reported by Hambsch, \emph{et al.}~ \cite{HAMBSCH1997347} (see Fig.~\ref{YA3}), the experimental post-neutron fragment mass distribution reported by Romano, \emph{et al.}~\cite{Romano} (see Fig.~\ref{fig:YApost}) and experimental post-neutron fragment mass distribution taken from \cite{MEIERBACHTOL201559} (see Fig.~\ref{YA2}), respectively. The blue solid line with dots represents the results for the default \cgmf{} input. The default \cgmf{} input is obtained by fitting the experimental pre-neutron fragment mass distribution reported by Hambsch~\emph{ et al.,} \cite{HAMBSCH1997347} to a three-Gaussian model (Eq.~\eqref{3Gaussian}, see parameters in Table.~\ref{Table2}). Notably, the distribution for the default \cgmf{} differs from the others because the mass resolution effects are removed in the deblurred distributions. However, differences are also observed in the deblurred distribution from Hambsch, \emph{et al.}~\cite{HAMBSCH1997347} compared to the other deblurred distributions.

Also, from Fig. \ref{CGMFdistribution}, it is clear that the pre- and post-neutron fragment mass distributions reported from the same experiment \cite{Romano} yield nearly identical deblurred distributions, as shown in Fig.~\ref{CGMFdistribution} by the solid and dot-dashed lines. In addition, comparing the results for inputs obtained by deblurring experimental pre-neutron FF mass distribution from \cite{HAMBSCH1997347} with the results for default \cgmf{} input is insightful, since here, the default \cgmf{} input is also obtained using data from~\cite{HAMBSCH1997347} as mentioned earlier.
Note that all \cgmf{} calculations discussed in this paper are based on the mass distributions shown in Fig.~\ref{CGMFdistribution}. To avoid repetition, we use the following abbreviations for the deblurred mass distributions inputs: YApre-Romano, YApre-Hambsch and YApost-Romano for mass distribution obtaining by deblurring the experimental pre-neutron mass distribution from \cite{Romano}, \cite{HAMBSCH1997347}, and  by deblurring the experimental post-neutron mass distribution from \cite{Romano}. We refer to the default \cgmf{} input simply as default.

\begin{figure}[hpt]
    \centering
    \includegraphics[width=1\linewidth]{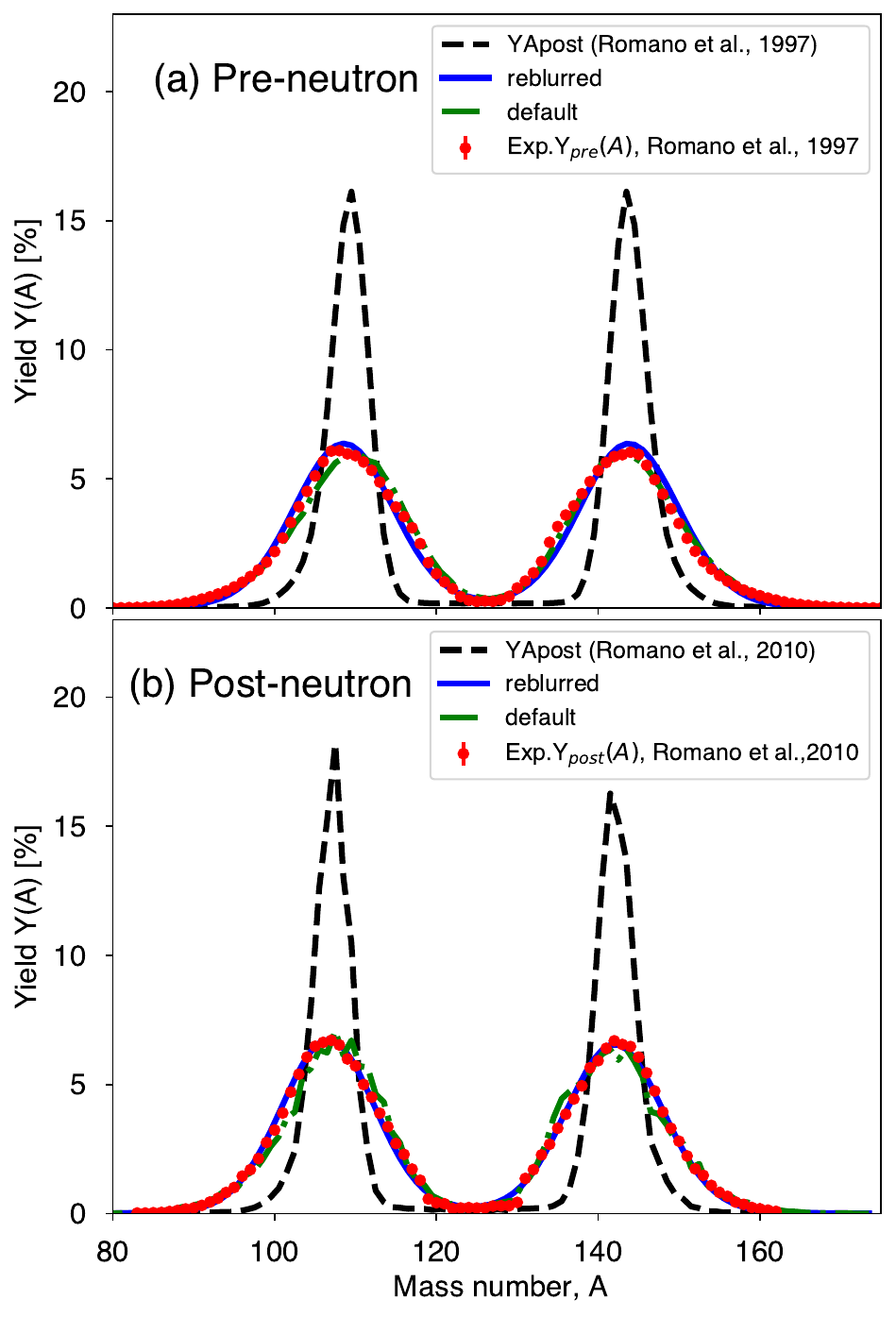}
    \caption{Fission Fragment mass distribution obtained from \cgmf{} for deblurring input, YApost-Romano for $^{252}$Cf(sf). (a) black dashed line  shows  calculation from \cgmf{}, blue solid line represents the reblurred distribution. The defauft \cgmf{} calculation and experimental $Y_{pre}(A)$ by Romano, \emph{et al.} \cite{Romano} are diplayed as green dash-dotted line and red dots, respectively. (b) We show post-neutron FF mass distribution caclulation from \cgmf{} by dashed line, blue solid line was obtained by reblurring the dashed line and green dot-dashed line displays default \cgmf{} post-neutron emission FF mass distribution, and red dots represents measured $Y_{post}(A)$  by Romano, \emph{et al.} \cite{Romano}. In both panels, reblurred and defauft distribution are in good agreement with measurements.}
    \label{CGMFdistribution1}
\end{figure}
\begin{figure}[hpt]
    \centering
    \includegraphics[width=1\linewidth]{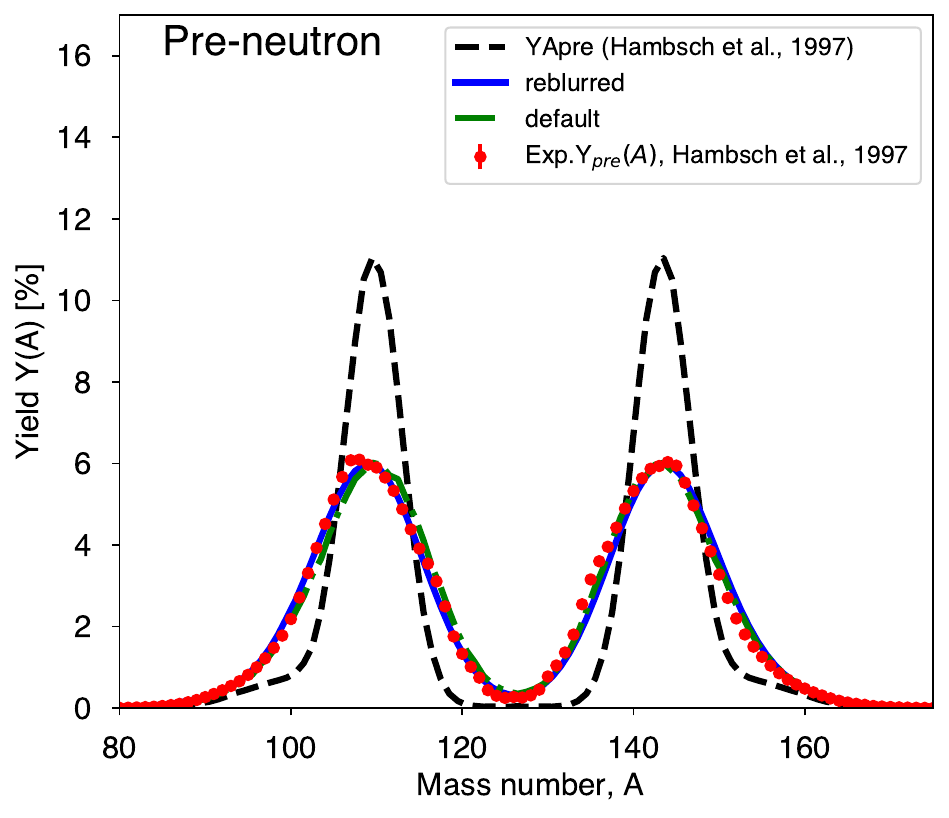}
    \caption{Comparison between reblurred pre-neutron emission FF mass distribution \cgmf{} calculations for deblurred input (YApre-Hambsch), default and measurements. Dashed line represents $Y_{pre}(A)$ from CGMF, solid blue and dot-dashed lines show reblurred $Y_{pre}(A)$ and default \cgmf{} $Y_{pre}(A)$, respectively. Additionally, we plotted experimental data: red dots shows $Y_{pre}(A)$ by Hambcsh, \emph{et al.} \cite{HAMBSCH1997347}. Clearly, reblurred and default distributions are in good agreement and they both reproduce measured $Y_{pre}(A)$.} 
    \label{CGMFpre}
\end{figure}

In Figs.~\ref{CGMFdistribution1} and~\ref{CGMFpre}, we show the $Y(A)$ distributions from \cgmf{} using deblurred inputs, indicated by the black dashed lines. These results are compared to experimental data from Romano \emph{et al.}~\cite{Romano} (Fig.~\ref{CGMFdistribution1}) and Hambsch~\emph{et al.}~\cite{HAMBSCH1997347} (Fig.~\ref{CGMFpre}). We also include the reblurred distributions, obtained by convolving the black dash lines with the TM, as well as the FF mass distribution from the default \cgmf{} calculation.

In Fig.~\ref{CGMFdistribution1}, we plot the $Y(A)$ distribution from \cgmf{} using the YApost-Romano input. Panel (a) shows the pre-neutron FF mass distributions: the black dashed line indicates the \cgmf{} $Y_{\text{pre}}(A)$, the blue solid line represents the reblurred version, and the experimental data from Romano \emph{et al.}~\cite{Romano} are shown as red dots. The default \cgmf{} distribution is represented by the green dash-dotted line. There is clear agreement between the reblurred and measured distributions. This is expected, as the reblurred $Y_{pre}(A)$ is derived from a \cgmf{} calculation using deblurred inputs—these inputs were themselves obtained by deblurring the experimental data from Romano~\emph{et al.}~\cite{Romano}.
In addition, Fig.~\ref{CGMFdistribution1}(b) displays the $Y_{post}(A)$ from \cgmf{} (black dashed line), its corresponding reblurred distribution (blue solid line), the default post-neutron \cgmf{} distribution (green dash-dotted line), and the measured post-neutron distribution from Romano~\emph{et al.}~\cite{Romano} (red dots). Again, we observe good agreement between the reblurred and measured distributions. Also, it is important to note that the TM used in reblurring $Y_{\text{pre}}(A)$ and in the calculation of $Y_{\text{pre}}(A)$ from \cgmf{} in Fig.~\ref{CGMFdistribution1}~(a) and (b) is the same as the one used to generate the YApostRomano inputs (see the parameter $\sigma_{\text{TM}}$ in Table~\ref{tab:my_label}). However, here we turned off the neutron multiplicity correction, since only resolution effects are needed for the reblurring.

It is worth noting that the difference observed at the light-fragment peak in the default \cgmf{} distribution, compared to the reblurred and measured distributions, is due to the fact that the default distribution was obtained by fitting the $Y_{\text{pre}}(A)$ data from Hambsch \emph{et al.}~\cite{HAMBSCH1997347} using the three-Gaussian model [Eq.~\eqref{3Gaussian}], whereas the experimental data shown here are from Romano \emph{et al.}~\cite{Romano}.

Similarly, we run \cgmf{} using the YApost-Hambsch input. In Fig.~\ref{CGMFpre}, the black dashed line represents the $Y_{\text{pre}}(A)$ from \cgmf{} using the YApre-Hambsch input. The blue solid and green dash-dotted lines correspond to the reblurred $Y_{\text{pre}}(A)$ and the default \cgmf{} $Y_{\text{pre}}(A)$, respectively. Additionally, the experimental data from Hambsch \emph{et al.}~\cite{HAMBSCH1997347} are shown as red stars. The reblurred and default distributions show good agreement with each other and successfully reproduce the measured $Y_{\text{pre}}(A)$. Such good agreement is expected, since both the deblurred and default \cgmf{} inputs are derived from the same measurements by Hambsch \emph{et al.}~\cite{HAMBSCH1997347}. Here, the TM used in reblurring is the same as the one used to generate YApre-Hambsch input.

In general, we observe consistent agreement between the reblurred and experimental mass distributions. Additionally, there is excellent agreement between the reblurred and default \cgmf{} distributions when both the default and deblurred inputs are derived from the same measurements. In this case, both the reblurred and default \cgmf{} results reproduce the experimental data well. Minor differences may arise from uncertainties associated with fitting the distributions using the three-Gaussian model. As mentioned earlier in this section, the current \cgmf{} implementation uses the best-fit parameters of the three-Gaussian model to generate the mass distribution input. The tests performed in this subsection indicate that, to directly compare the fission fragment (FF) mass distribution results from \cgmf{} (using deblurred inputs) with experimental data, the calculated distributions must first be reblurred.

\subsection{Neutron and gamma-ray observables from \cgmf{} calculations}

To compare the results for default \cgmf{} inputs against inputs obtained by deblurring method, we calculate prompt neutron and $\gamma$-ray observables using \cgmf{}, including neutron multiplicity, the prompt fission neutron spectrum (PFNS), and the prompt fission $\gamma$-ray spectrum (PFGS), for the spontaneous fission of $^{252}$Cf. Here, for those observables, we present the results for four different fragment mass distribution inputs discussed earlier in the text. Three of these inputs were obtained using the deblurring method, and one corresponds to the \cgmf{} default input. 

It is worth noting that it would interesting to reblur other mass dependence observables, such as $ \Bar{\nu} (A)$ and total kinetic energy TKE(A) calculated from \cgmf{} using deblurred inputs, in order to account for experimental resolution effects, similar to what was done in the previous subsection for $Y(A)$. However, the infrastructure for performing such calculations is not yet available, and we reserve this for future work. Therefore, in this subsection, we focus only on comparing the \cgmf{} results obtained using deblurred versus default inputs.

First, we calculate the average prompt neutron multiplicity, $\overline{\nu}$ for different inputs, and obtain the following results: 3.751 (default), 3.705 (YApre-Romano), 3.753 (YApost-Romano), and 3.760 (YApre-Hambsch). These values are generally in good agreement with the standard value of $3.759 \pm 0.0048$, as reported in Ref.~\cite{standard}, except for the one obtained using YApre-Romano.

\begin{figure}[hpt]
    \centering
    \includegraphics[width=1\linewidth]{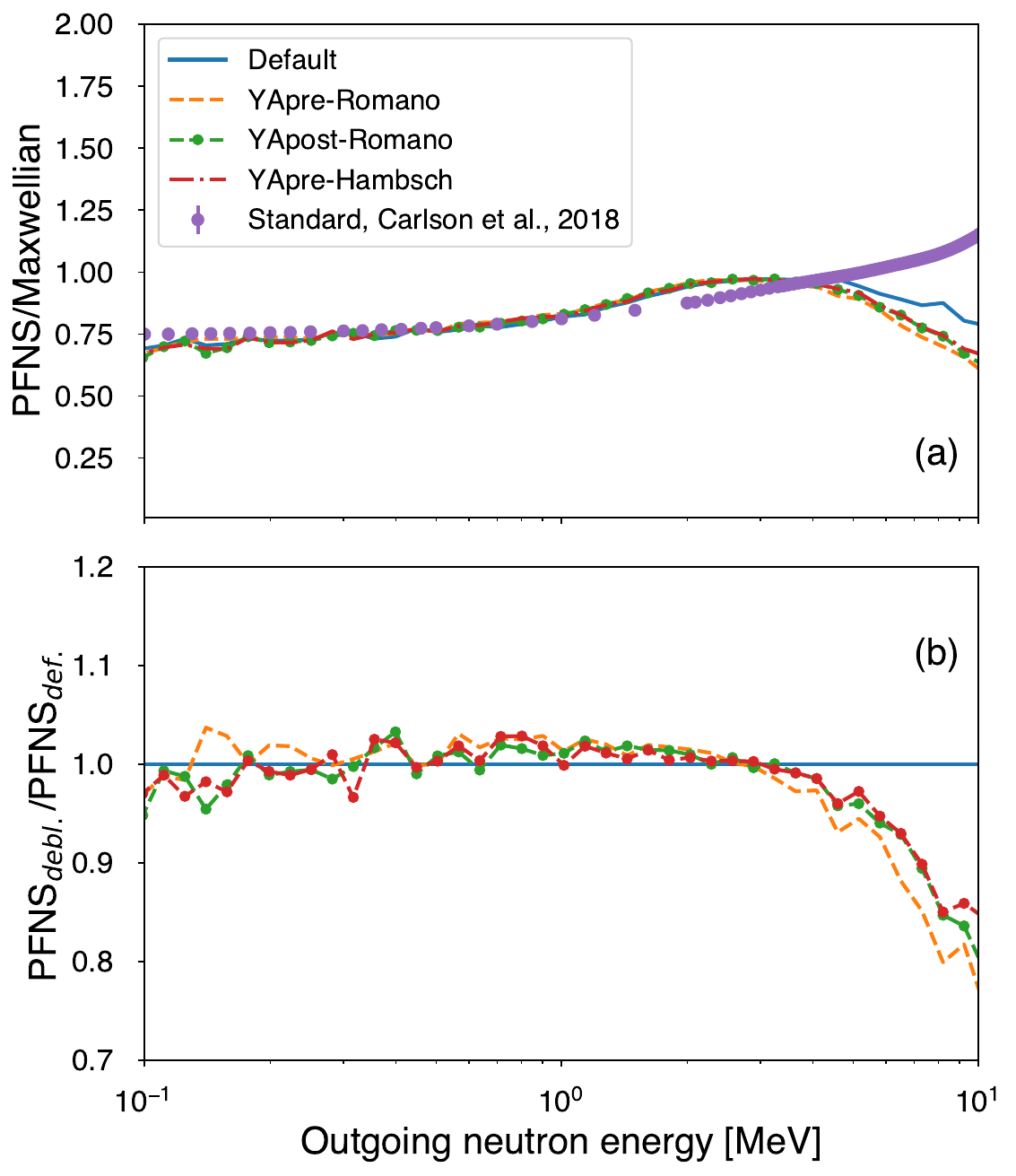}
    \caption{(a) The calculated and evaluated prompt fission neutron spectra for the spontaneous fission of $^{252}$Cf(sf) are shown as ratios to a Maxwellian spectrum at temperature $kT=1.32$ MeV. (b) The various \cgmf{} PFNS calculations are presented as ratios to the result obtained using the default input mass distribution.}
    \label{PFNS}
\end{figure}

Then, we calculate the PFNS for the default \cgmf{} and deblurring inputs and compare the results with experimental measurements. Shown in Figure~\ref{PFNS}(a) are the results of the PFNS as a ratio to a Maxwellian with temperature parameter, $kT=$1.32 MeV, for the default \cgmf{} input (represented by the solid line) alongside those for the deblurring inputs: YApre-Romano (dashed line), YApost-Romano (dashed line with dots), and YApre-Hambsch (dot dashed line). The evaluated PFNS for $^{252}$Cf(sf) reported by Carlson, \emph{et al.,}~\cite{standard}, is shown as filled circles.  We observe the calculated PFNS reproduce  the evaluated PFNS in the outgoing neutron energy range up to 5 MeV. However, in the high-energy tail, a significant difference between the calculated and experimental PFNS is observed. This discrepancy is a well-known challenge for Hauser-Feshbach fission fragment decay models. Also, the ongoing work by Neudecker \emph{et al.}~\cite{Denise} on the re-evaluation of the standard PFNS shows that their preliminary results are very close to the existing standard PFNS. Looking ahead, the developments of more accurate physics models, such as level densities for neutron-rich nuclei, in \cgmf{} are expected to shed further light on this discrepancy.

To better observe the difference between the results from the deblurring inputs and the default \cgmf{}, we plot the ratio between them in panel (b) of Fig. \ref{PFNS}. It can be seen that there is no notable difference between the calculations in the low outgoing neutron energy range, as the ratio is approximately equal to 1. However, a large deviation of the ratio from unity ($\sim$ 20\%) is observed in the high outgoing neutron energy region; in that region, the deblurred Y(A) distributions soften the PFNS. Our analysis indicates that high outgoing energy neutrons are predominantly emitted by fragments with masses near 120~. Since the deblurred Y(A) exhibits reduced yields around this mass compared to the default Y(A), fewer such neutrons are emitted, contributing to the observed softening in PFNS. However, other \cgmf{} inputs such as the fragments' kinetic energy and the discrete level density of the fission fragments can also contribute to the shape of the PFNS and could compensate for the softening introduced here. 

In addition, we investigate the impact of deblurring inputs on prompt fission $\gamma$-ray observables. 
\begin{figure}[hpt]
    \centering
    \includegraphics[width=1\linewidth]{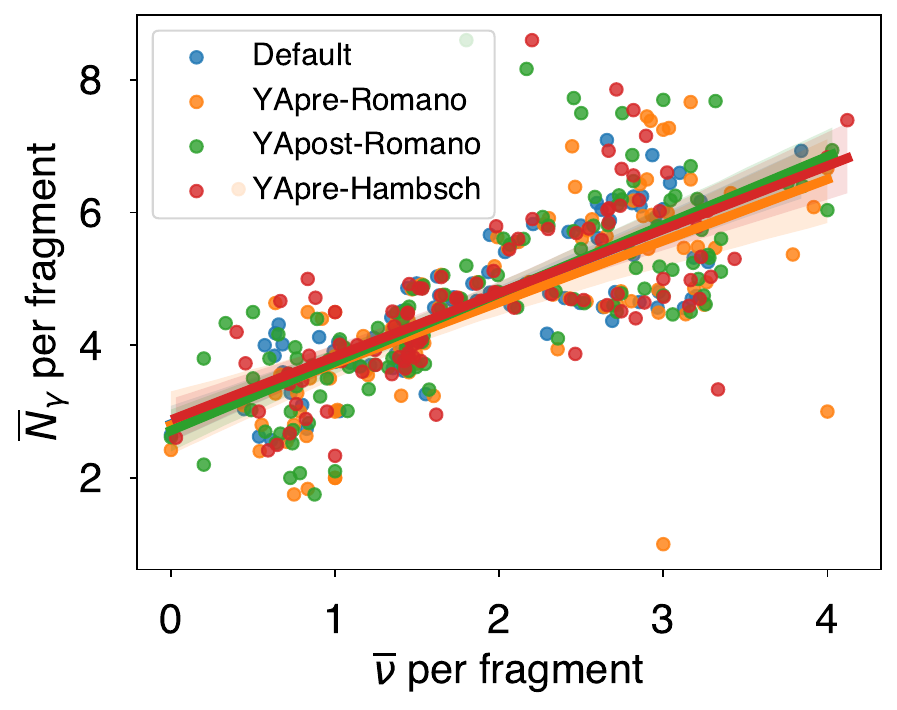}
    \caption{Correlations between the calculated average $\gamma$-ray multiplicity ($\overline{N}_\gamma$) and neutron multiplicity ($\overline{\nu}$) per fragment for the spontaneous fission of $^{252}$Cf. Different colors correspond to calculations from different inputs as shown in the legend.}
    \label{nugcorr}
\end{figure}
\begin{figure}[hpt]
    \centering
    \includegraphics[width=1\linewidth]{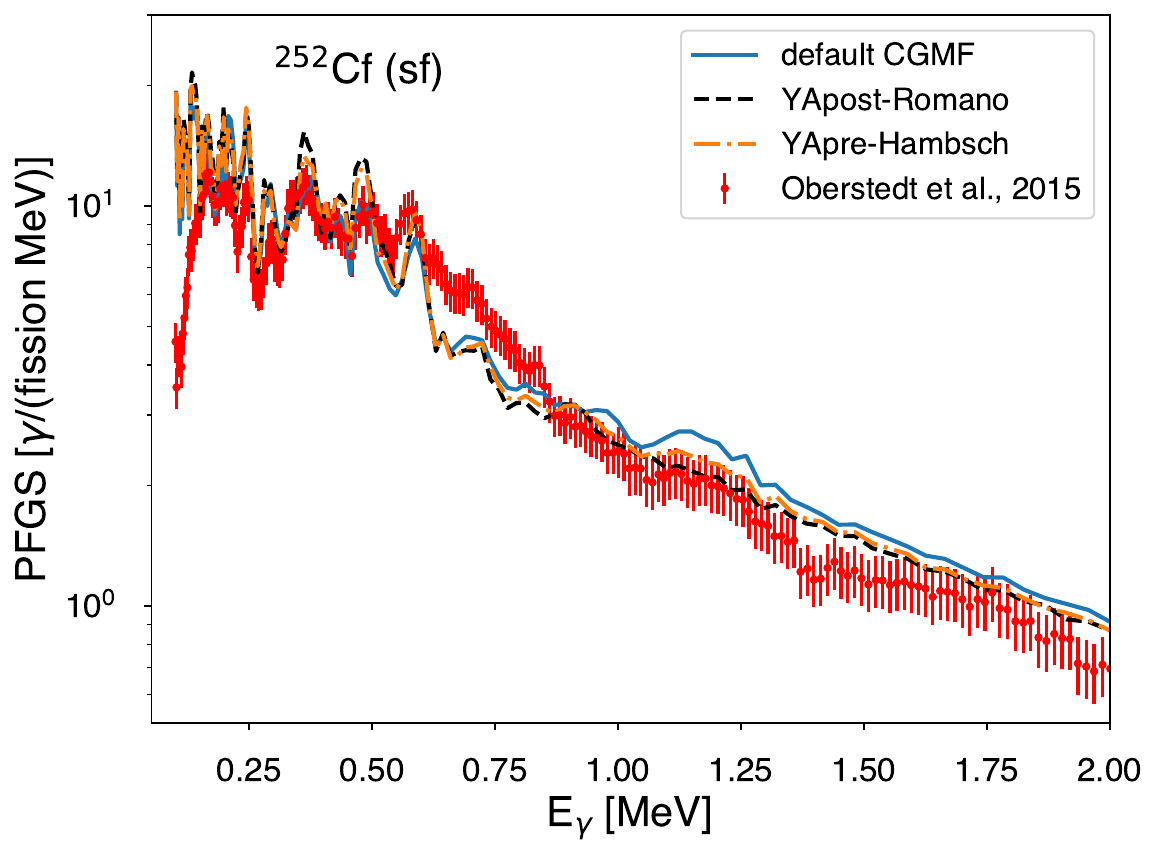}
    \caption{ The prompt fission $\gamma$-ray spectrum (PFGS) as a function of outgoing $\gamma$-ray energy for the spontaneous fission of $^{252}$Cf. The solid line shows the results obtained from \cgmf{} calculations using default \cgmf{} inputs. The dashed and dot-dashed lines represent the results from \cgmf{} calculations with deblurred inputs, YApost-Romano and YApre-Hambsch, respectively. Dots display the experimental PFGS reported by Oberstedt, \emph{et al.}~\cite{oberstedt}. The calculations qualitatively reproduce experimental data.}
    \label{pfgs}
\end{figure}
Here, we first examine the correlation between the calculated average $\gamma$ multiplicity ($\overline{N}_{\gamma}$) and average neutron multiplicity ($\overline{\nu}$) per fragment, plotted in Fig.~\ref{nugcorr}. For all inputs to \cgmf{}, there is a clear positive correlation between $\overline{N}_{\gamma}$ and $\overline{\nu}$, which means that as the average number of neutrons emitted per fragment increases, the average number of $\gamma$ rays emitted per fragment also increases. This correlation was also reported in Ref.~\cite{wang_corr}. The dots of different colors correspond to different $Y_{pre}(A)$ inputs into \cgmf{}. Overall the correlation trends are preserved, regardless of the Y(A) used.

\begin{figure*}
    \centering
    \includegraphics[width=1\linewidth]{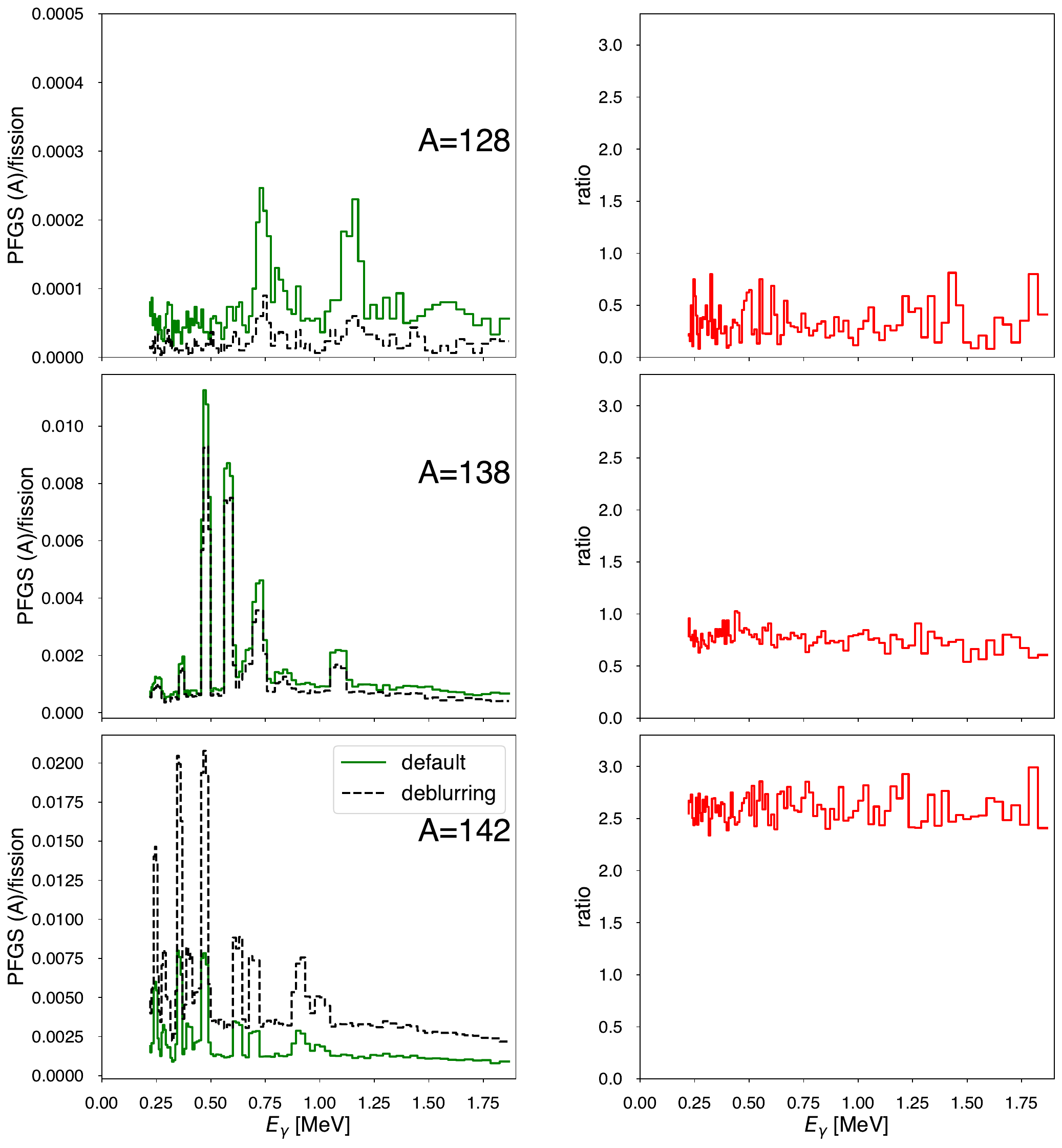}
    \caption{The PFGS for selected fission fragments (see the panel labels). The peaks in each fragment PFGS represent the contribution of that fragment to the total PFGS shown in Fig.~\ref{pfgs}. The solid lines represent the results obtained with the default \cgmf{} inputs and the dotted lines correspond to the results obtained with deblurred FF mass distribution input (YApre-Hambsch). On the right, we show the corresponding ratio of deblurring to default PFGS.}
    \label{Apfgs}
\end{figure*}

Additionally, we calculate the PFGS as a function of outgoing $\gamma$-ray energy ($E_\gamma$), as shown in Fig.~\ref{pfgs}. In the figure, the solid line represents the PFGS calculation with the default \cgmf{} input, and the dashed and dash-dotted lines show the results from \cgmf{} calculations for $Y_{pre}(A)$ inputs from deblurring, YApost-Romano and YApre-Hambsch inputs, respectively. The red dots show the experimental PFGS values from Oberstedt, \emph{et al.} \cite{PFGS}. The observed peaks in the low-energy part of the spectrum correspond to the excitation of collective discrete states in the fission fragments. The energies and decay data of those states are input to \cgmf{}, and taken mostly from the RIPL-3 database~\cite{RIPL3}, but their excitation depends on the decay paths calculated within the code and depends on various physics models. Calculations for both default and deblurred inputs to \cgmf{} qualitatively reproduce the experimental data. However, we can also observe discrepancies in the PFGS for different results over the $E_\gamma$ ranged plotted here; for $E_\gamma \leq 0.5$ MeV, generally the results from deblurring method overpredict the experimental PFGS, while at high $\gamma$-ray energies, $E_\gamma \ge 0.75$ MeV, the deblurred results are within the uncertainties of the experimental data and are better than default \cgmf{} calculations. While \cgmf{} does not completely agree with the data, it reproduces nicely most if not all the observed structures. The amplitude of each peak is also going to be dictated by some experimental setup conditions, e.g., coincidence time and energy threshold. Interestingly, for the $0.6 \leq E_\gamma \le 0.8$ MeV range, both deblurred and default \cgmf{} predictions are lower compared to experimental data in that range, and this is also observed in data set from ~\cite{PFGS,Chyzh,VV}.  

\begin{figure}
    \centering
    \includegraphics[width=0.9\linewidth]{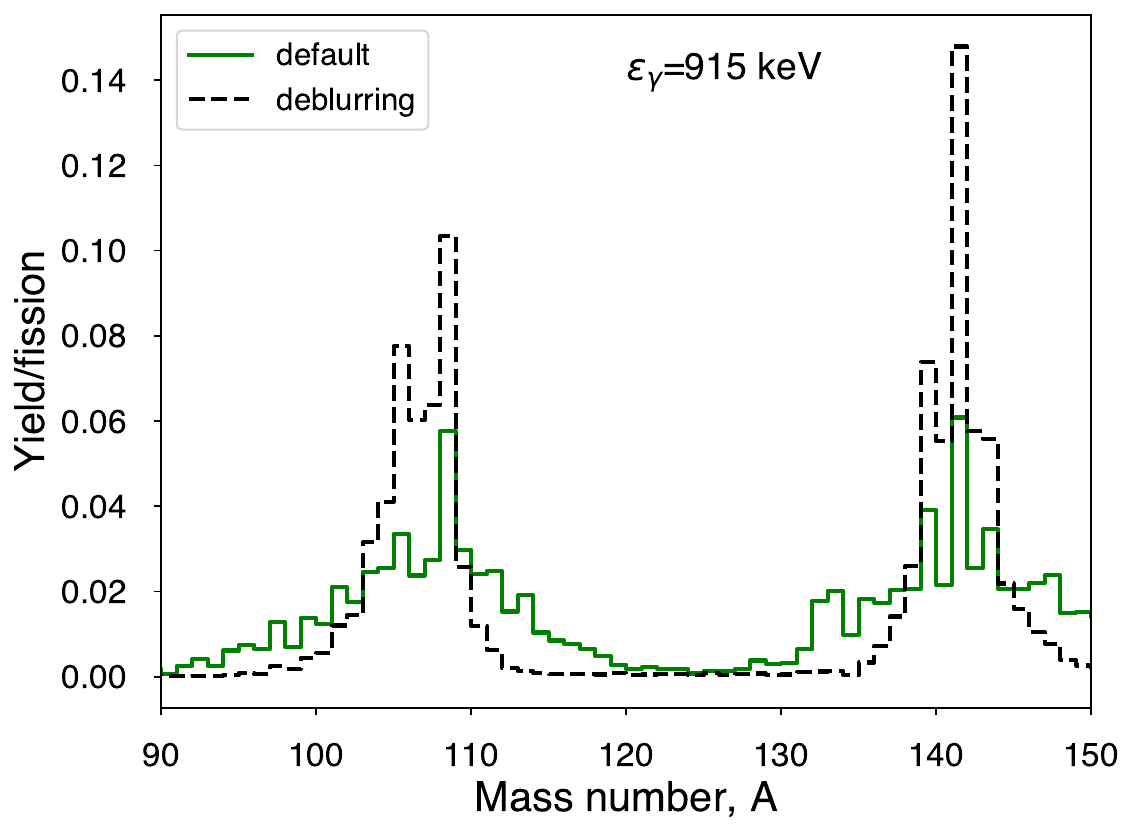}
    \caption{Fragment distributions at E$=0.915$ MeV, the vertical axis represents yield. The results for input from the deblurring method (YApre-Hambsch) are shown as a dashed line, and from \cgmf{} default inputs are shown as a solid line. We observe a strong peak around fragment mass, $A=$142. } 
    \label{e915}
\end{figure}

To begin to investigate these discrepancies, we plot the contributions of individual fission fragment to the PFGS. Here, we select fragments based on their mass number. As an example, Fig.~\ref{Apfgs} shows the PFGS for selected fragment masses: $A = 128$, 138, and 142. These fragments correspond, respectively, to masses near symmetry, near the crossing point of the deblurred and default mass distributions, and near the peak of the mass distribution. We choose these fragment masses because they provide representative information about differences in the PFGS between deblurred and default results and offer a good characterization of the overall mass distribution. The difference between the deblurred and default PFGS is clear, especially fragment near the peak, $A =142$. The deblurred PFGS shows higher yields compared to the default, as shown by their ratio in the right column of the figure. If spectroscopy could be performed on fission fragments of this mass and other fragments around this peak, we could gain insights into the pre-neutron emission mass distributions. Each panel here represents the ratio of the deblurred to default \cgmf{} PFGS for the mass in the corresponding panel of the left column. The structures observed in the ratio indicate how much the two results differ at each $E_\gamma$. But for all mass numbers, the position of the peaks is consistent for both sets of results, which implies the consistency of both methods in providing physics information about PFGS. Also, we observe that  mass $A=$142 has a distinguishable structure around $E_\gamma= 0.9$ MeV compared to other fragment masses. This suggest that this mass contributes much more to the broad peak of the PFGS, Fig.~\ref{pfgs}, around that $E_\gamma$. To verify that we plot in Fig.~\ref{e915}, the fragment distribution for $E_\gamma=$0.915 MeV, and it is seen that the peak around $A=$142 has higher amplitude compare to other peaks, especially for deblurred results.

In general, we observe some differences between the PFGS from the deblurring method and the default \cgmf{}, particularly for low outgoing $\gamma$-ray energies. However, the deblurring method still provides reasonable results as it reproduces the features (physical information) observed in the experimental PFGS. Additionally, in the range of $E_\gamma = 0.5\--0.8$ MeV, we observe a discrepancy between the calculated and experimental PFGS, which could be due to a missing discrete level in the input file. We are currently investigating this feature.\\

\section{summary
\label{sec:summary}}
In this paper, we demonstrate the use of the Richardson-Lucy deblurring algorithm, which is commonly applied in optics and has recently been introduced to the field of nuclear physics. We use the algorithm to extract pre-neutron FF mass distributions from measured post-emission neutron FF mass distributions and to remove mass resolution from the reported pre-neutron-emission FF mass distribution. We considered the experimental measurements  of spontaneous fission of $^{252}$Cf.

Using different experimental data sets, we demonstrate that the deblurring method significantly reduces the experimental mass resolution effects in the measured distribution.
Additionally, when the deblurring is applied to pre-neutron FF mass distributions obtained using other methods (but still affected by mass resolution effects), it reduces those effects and produces distributions similar to those obtained when deblurring is applied directly to the measured post-neutron FF mass distribution.

Moreover, the pre-neutron FF mass distribution is an important input for Monte Carlo simulations used in nuclear fission codes, such as the \cgmf{} code. We used the deblurred distribution as input to \cgmf{} and compared the resulting prompt fission observables with those obtained from default \cgmf{} calculations. In particular, we examined the correlation between prompt fission neutrons and $\gamma$ rays and observed good agreement between the results from both methods. However, large differences in $\gamma$-ray observables are observed between the deblurring and default \cgmf{} results, suggesting that these observables could be used to gain insights into the pre-neutron emission mass distributions.

Finally, the deblurring method offers a promising approach to extracting pre-neutron emission fission fragment mass distributions from measured data and could be extended to other fission fragment properties. In the future, we are considering using the RL algorithm with more realistic detector response matrices, for example, the response matrix of SPIDER, a new instrument for measuring fission fragments developed at Los Alamos National Laboratory, to gain deeper insight into the pre-neutron emission fission fragment mass distribution. Also, we plan on applying this deblurring technique as a novel unfolding method to extract fission fragment properties from prompt fission neutron and $\gamma$-ray spectra, especially in very low-energy and high-energy regimes where spectra fluctuate significantly. 

\acknowledgments
This work was performed under the auspice of the U.S. Department of Energy by Los Alamos National Laboratory under Contract 89233218CNA000001.  This work was supported by the Advanced Simulation and Computing (ASC) program at Los Alamos National Laboratory.

\newpage
\bibliographystyle{apsrev}
\bibliography{ref}

\clearpage
\appendix
\section{Testing the deblurring method } \label{testsec}
\begin{figure*}
    \centering
    \includegraphics[width=1\linewidth]{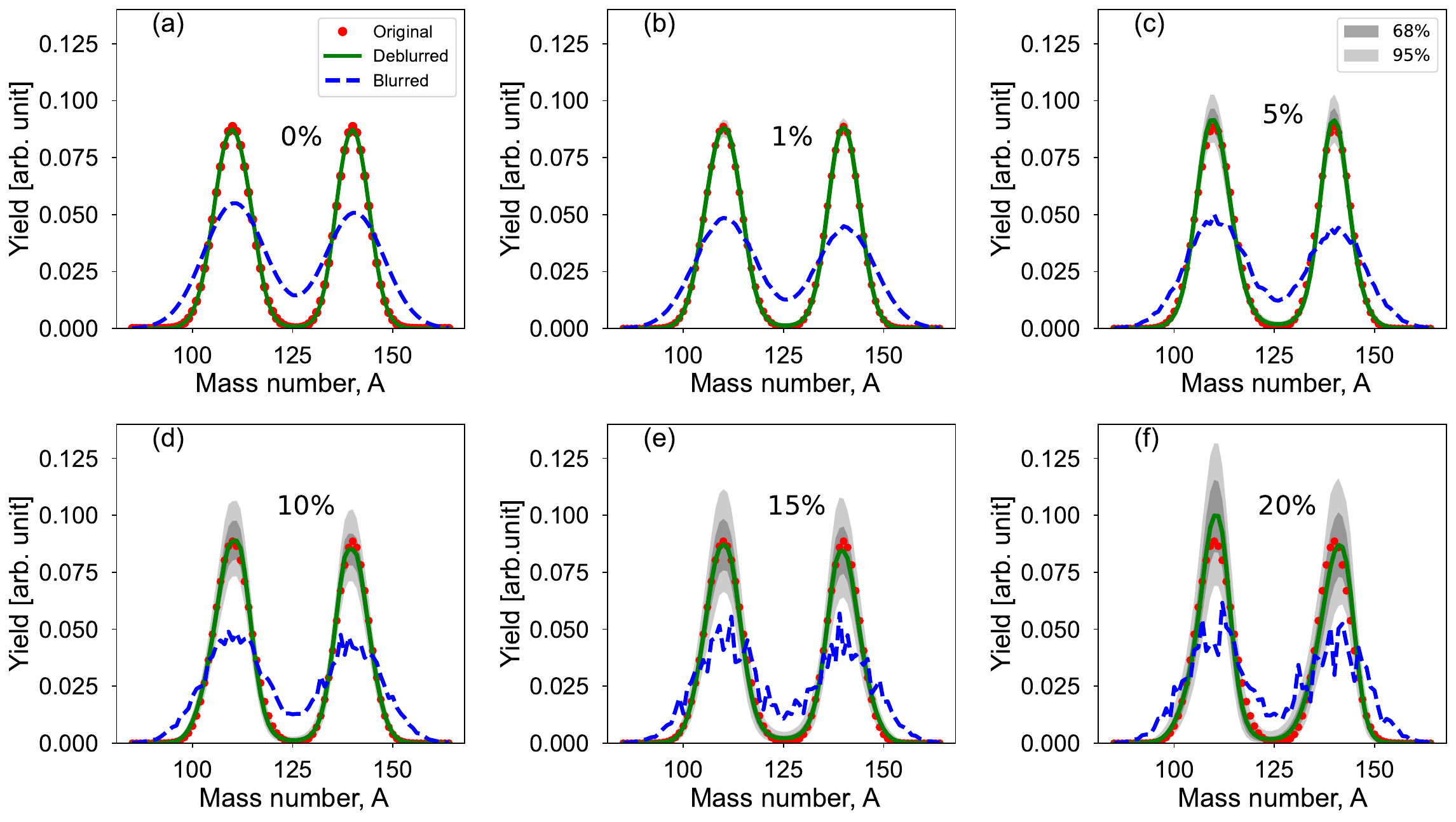}
    \caption{Deblurring of the two-Gaussian model with varying levels of noise added to the blurred distribution. Panel (a) shows the case with no added noise; panels (b)–(f) correspond to added noise levels of 1\%, 5\%, 10\%, 15\%, and 20\%, respectively. The dots represent the original distribution defined by Eq.~\eqref{two-gau}, while the dashed line shows the blurred distribution. The dark and light gray bands indicate the 1$\sigma$ and 2$\sigma$ uncertainties, respectively, resulting from deblurring distribution with error sampling. We generated 1000 samples from this error sampling. The solid line represents the deblurred result; in panels (b)–(f), it corresponds to the average of the 1000 deblurred samples. Overall, the original distribution is successfully restored in all cases.}
    \label{test1}
\end{figure*}

Here, we perform tests on our deblurring method as applied to known distributions. First, we consider cases where the errors are negligible, followed by scenarios in which random errors are introduced. In line with our interest in fission fragment mass distributions which typically exhibit a Gaussian shape, the test distribution, $\mathcal{D}(A)$, is modeled using a two-Gaussian model
\begin{eqnarray}
    \mathcal{D}(A)= \sum _{i}W_i\frac{1}{\sqrt{2\pi\sigma_i}}\exp{\Big[-0.5\big (\frac{A-\mu_i}{\sigma_i}\big)^2}\Big],
    \label{two-gau}
\end{eqnarray}
where $i = 1, 2$, $W_i$ is the weight, and $\sigma_i$ and $\mu_i$ are the standard deviation and mean of the Gaussian distributions, respectively, and $A$ represents the mass number. In the illustrative case used here, we take $\mu_{1,2} = 110$ and $140$, and $\sigma_{1,2} = 4.5$ and $4.0$. The value of $A$ ranges from 85 to 165 (see Fig.~\ref{test1}(a)). For simplicity, we set $W_i = 1$, which implies that the function $\mathcal{D}(A)$ is normalized to 2.

First, we take the modeled distribution $\mathcal{D}$ and multiply it with TM, Eq.~\eqref{TM} of $\sigma _{TM}=6.2$, to obtain a blurred distribution, $d(A)$. Both $\mathcal{D}$ and $d$ represent distributions in the limit of high statistics. The blurred distribution $d$ is then used as input to the Richardson–Lucy (RL) deblurring algorithm, as described in Eqs.\eqref{eq:RL} and \eqref{RLr}, to reconstruct a deblurred distribution, estimate of the original distribution. All three distributions $\mathcal{D}$ (original), $d$ (blurred), and the deblurred result are shown in Fig.\ref{test1}(a). Notably, the deblurred distribution overlaps almost perfectly with the original $\mathcal{D}$, demonstrating that the algorithm successfully recovers the underlying features of the original distribution. This consistency holds regardless of the specific shape of $\mathcal{D}$ used in the simulation.

To test the robustness of the method under more realistic conditions, we introduce random fluctuations by adding a Gaussian noise term $\mathcal{N}(0, \epsilon(A))$ to $d$, resulting in a noisy distribution $d^* = d + \mathcal{N}(0, \epsilon(A))$. Here, $\epsilon(A)$ represents the standard deviation of the noise and is taken as a percentage of $d(A)$. We consider several noise levels: $\epsilon(A) = 1\%$, 5$\%$, 10$\%$, 15$\%$, and 20$\%$ of the blurred distribution. The RL algorithm is then applied to each noisy dataset $d^*$ to reconstruct an estimate of the original distribution $\mathcal{D}$. The noisy distributions correspond to a distribution with limited statistics.

 In Figure~\ref{test1}, we compare the original and deblurred distributions when statistical fluctuations are introduced into the blurred distribution. The blurred distribution is also shown for reference. Points represent the original distribution, while solid and dashed lines correspond to the deblurred and blurred distributions, respectively. Panel (a) shows the results with no added noise; panels (b) through (f) correspond to added noise levels of 1\%, 5\%, 10\%, 15\%, and 20\%, respectively.

To gain further insight into error propagation, we generate an ensemble of noisy distributions by sampling each point of $d^*$ according to a normal distribution, $\mathcal{N}(d^*(A), \epsilon(A))$, which ensures that the generated samples remain within the noise level considered for $d^*$. We generate 1000 such samples, this number is arbitrary and can be adjusted. Each sampled distribution is then deblurred using the RL algorithm, resulting in a set of estimated distributions for $\mathcal{D}$. From this ensemble, we compute the 1$\sigma$ (68$\%$) and 2$\sigma$ (95$\%$) confidence intervals, shown as dark and light gray bands, respectively, in Figure~\ref{test1}.

For panels (b) through (f), the solid green line represents the mean of the deblurred distributions across the ensemble. This serves as a reference to assess the average performance of the RL algorithm in reconstructing the original distribution from noisy input. As expected, the fluctuations in the blurred distribution grow with increasing noise level, becoming particularly significant at the 10$\%$ level and beyond. Consequently, the uncertainty bands in the deblurred distributions also widen. Nevertheless, the RL algorithm successfully reproduces the original distribution on average for all noise levels considered. However, in the case of 20$\%$ noise, a noticeable shift on the x-axis appears between the deblurred and original distributions. This discrepancy may arise from the larger, randomly distributed errors, which can introduce significant local fluctuations in the blurred input, especially in regions with low values of $d$, ultimately impacting the RL algorithm’s reconstruction accuracy in those regions. 

In real experiments, a 20\% noise level may be considered high. However, if the RL algorithm performs well under such conditions, it ensures that it will also work reliably with real experimental measurements. It is also known that the RL algorithm tends to struggle in regions of the distribution with very low values, for example, where $\mathcal{D}_i \leq 10^{-4}$ in our test case. Nonetheless, the algorithm successfully reconstructs the overall shape of the original distribution, as shown in Fig.~\ref{test1}.
It is important to mention that to prevent error amplification during the iterative process, we apply TV regularization, the method described in Ref.~\cite{NZABAHIMANA2023138247} to RL algorithm for the blurred distribution with a high error level (i.e., error greater than 10\%), using a regularization factor of 0.001. However, we found that for distributions with a small error level, applying regularization is unnecessary.


\begin{figure*}
    \centering
    \includegraphics[width=1\linewidth]{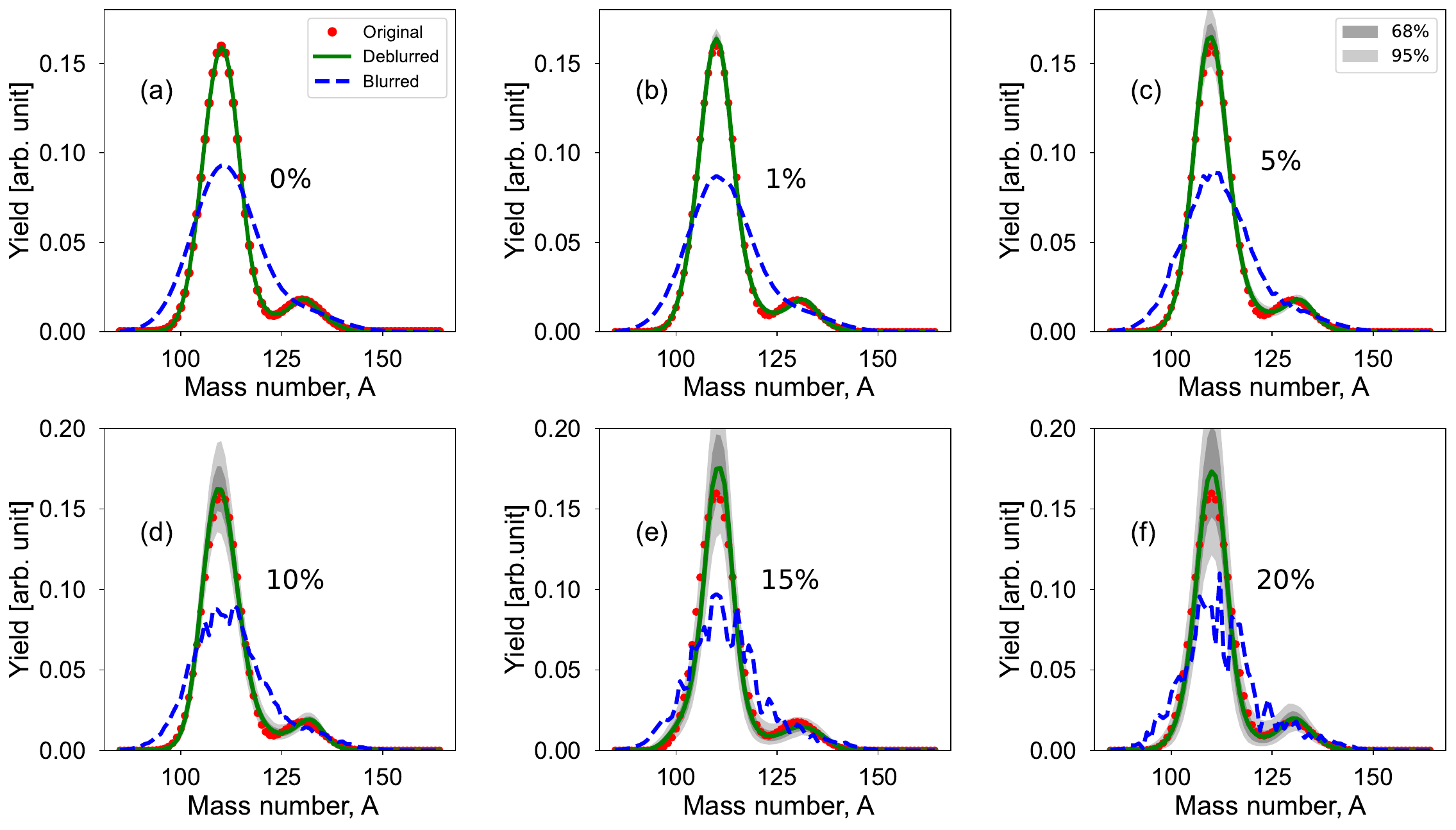}
    \caption{
    Similar to Fig.~\ref{test1}, but here the original distribution is generated from a two-Gaussian function as defined in Eq.~\eqref{two-gau}, with parameters $W_{1,2} = 1.8,\, 0.2$, $\sigma_{1,2} = 4.5,\, 5$, and $\mu_{1,2} = 110,\, 130$.
    }
    \label{test2}
\end{figure*}

We present an additional test case to evaluate the performance of the RL algorithm. We consider a two-Gaussian function as described in Eq.~\eqref{two-gau}, with the parameters $W_{1,2} = 1.8,\, 0.2$, $\sigma_{1,2} = 4.5,\, 5$ , and $\mu_{1,2} = 110,\, 130$ . The indices 1 and 2 refer to the parameters of the first and second Gaussian components, respectively. This function, shown as red points in Fig.~\ref{test2}, is treated as the original distribution.

As discussed in Section~\ref{testsec}, the blurred distribution is obtained by convolving the original distribution with the transfer matrix (TM), defined in Eq.~\eqref{TM}. In this example, we use $\sigma_{\text{TM}} = 6.2$ . The resulting blurred distributions are shown as dashed blue lines. Panel (a) represents the case with negligible error, while panels (b) through (f) include added random Gaussian noise at levels of 1\%, 5\%, 10\%, 15\%, and 20\%, respectively.

Our goal is to apply the RL algorithm to these blurred distributions, with and without added noise, to reconstruct the original signal. In the figure, the green solid lines show the reconstructed (deblurred) distributions. The dark and light gray bands correspond to $1\sigma$ (68\%) and $2\sigma$ (95\%) confidence intervals, respectively, obtained from sampling points in the blurred distribution (See the discussion in the text regarding Fig.~\ref{test1}).

This example was specifically chosen to test the RL algorithm’s ability to recover smaller features such as low-amplitude peaks that are not visible in the blurred distribution. This scenario simulates realistic challenges that may arise in actual experimental data.

\end{document}